\documentclass[twocolumn,english,APS, PRA, Eqsecnum]{revtex4-2}
\usepackage[T1]{fontenc}
\usepackage[latin9]{inputenc}
\usepackage{amsmath}
\usepackage{amssymb}
\usepackage{graphicx}
\usepackage{wasysym}
\usepackage{babel}
\begin{document}
\title{Simulating macroscopic quantum correlations in linear networks}
\author{A. Dellios, Peter D. Drummond, Bogdan Opanchuk, Run Yan Teh, and Margaret
D. Reid}
\affiliation{Swinburne University of Technology, Melbourne 3122, Australia}
\begin{abstract}
Many developing quantum technologies make use of quantum networks
of different types. Even linear quantum networks are nontrivial, as
the output photon distributions can be exponentially complex. Despite
this, they can still be computationally simulated. The methods used
are transformations into equivalent phase-space representations, which
can then be treated probabilistically. This provides an exceptionally
useful tool for the prediction and validation of experimental results,
including decoherence. As well as experiments in Gaussian boson sampling,
which are intended to demonstrate quantum computational advantage,
these methods are applicable to other types of entangled linear quantum
networks as well. This paper provides a tutorial and review of work
in this area, to explain quantum phase-space techniques using the
positive-P and Wigner distributions.
\end{abstract}
\maketitle

\section{Introduction}

There is a widespread interest in current physics in the immense variety
of macroscopic quantum correlations that can occur in physical systems
that are not in thermal equilibrium. Examples include quantum networks,
which are a growing area of modern quantum technology. These may operate
either in a transient mode \citep{zhong2020quantum}, or in the steady-state
\citep{honjo2021100}. In contrast to traditional condensed matter
physics, these are typically very far from being homogeneous. Such
networks are qualitatively different to small systems with low excitation
and are also distinct from thermal equilibrium studies in many-body
systems \citep{Book_CondensedMatter_integrals}, since external inputs
drive them into non-thermal-equilibrium states.

Systems of this type are both highly challenging and fascinating to
modern physicists from several points of view. The challenging aspect
is that they are hard to solve. Macroscopic quantum networks have
an exponentially large Hilbert space \citep{AaronsonArkhipov2013LV}.
This generally rules out exact orthogonal basis methods, for example
by expanding in the usual basis of number states. Analytic solutions
for the density matrix are rare. Even these are not always as useful
as one might expect, due to difficulties in computationally evaluating
the solutions except as averages over many possible implementations
\citep{drummond2016scaling}.

The fascination of these systems is that they are an unknown territory.
Sending probes to Mars is challenging, but even on Mars, telescopes
give us an idea of what to expect. Macroscopic quantum networks, because
of their exponential complexity and lack of analytic solutions, have
few scientific precedents. As well as giving rise to macroscopic Bell
violations, such networks introduce the issue of computational complexity,
which is simply that some calculations may take enormously long times.
Quantum correlations occur that may not cause a Bell violation, but
are still classically ruled out in a more subtle way. Some correlations
are computationally prohibited classically.

In addition to their fundamental interest, there is an increasingly
important potential in quantum technology. Some commonly researched
applications are to novel types of sensors \citep{Caves_PRD1981,McCuller2020PhysRevLett.124.171102}
or metrological devices \citep{Motes2015_PRL114,OlsonLinearOptQM,Su2017Multiphoton}.
Additionally, there are new types of quantum computers, whose purpose
is to perform a task that is thought to be impossible using classical
means \citep{boixo2018characterizing}. These include Gaussian boson
samplers \citep{AaronsonArkhipov:2011} and the coherent Ising machine
(CIM) \citep{Marandi_CIM_Nature2014}, which have now been scaled
to larger sizes than conventional gate-based quantum computers \citep{zhongPhaseprogrammableGaussianBoson2021,honjo2021100}.
The reason for this is that they have a much simpler design than conventional
quantum logic gate approaches. In the simplest cases they have only
linear coupling \citep{Aaronson2011} or just one type of nonlinearity
and no logic gates \citep{yamamoto2017coherent}. In a similar way,
reduced instruction set classical computers \citep{tanenbaum1978implications}
are often more scalable than older designs with large instruction
sets.

There is an ongoing debate about the advantages and limitations of
such devices, as well as the potential for verifying their behavior,
given the essential Hilbert space complexity. This leads to a distinction
between types of classical simulation, which we explain. While direct
simulations of every counting event are possible for small networks
\citep{neville2017classical,gupt2020classical,li2020benchmarking,quesada2020exact},
these rapidly become impossible at larger sizes, and may take billions
of years \citep{zhong2020quantum}. Despite this, one may still be
able to use phase-space methods to simulate photon-counting probabilities,
which allows verification of important network characteristics \citep{opanchuk2018simulating,drummond2020initial,drummond2021simulating}.
Such results are not restricted to low-order correlations, and can
test for decoherence and high order correlations.

Because this is a large and rapidly growing field, only a few aspects
of this work can be covered here. We treat linear networks, and nonlinear
cases such as the CIM are given elsewhere \citep{kiesewetter2021weighted}.
It is hoped that by giving the basics of this work, the reader may
be motivated to investigate further. The main point emphasized is
that even with growing network complexity, there are theoretical tools
available to analyze them. One cannot make predictions to $12$ decimals.
However, one can still compare models with experimental measurements,
which is essential both to test the theory and to improve the experiments.

\section{Phase-space simulations}

Why are quantum networks theoretically challenging?

The main feature of a quantum network is the number of modes involved.
This leads to exponential growth in the Hilbert space dimension. As
an illustration, if one has $100$ qubits, which have a basis of spin
up and down, there are $d=2^{100}\approx10^{30}$ quantum states.
The systems treated here comprise $M$ modes labelled $j=1,..,M$,
and with boson annihilation operators $\hat{a}_{j}$. Recent experiments
have already employed $M>100$ modes \citep{zhongPhaseprogrammableGaussianBoson2021}
in a linear network. Bosonic modes, with more than two states, have
an even faster growth in dimension.

To store the quantum amplitudes of even $10^{30}$ states with 15
decimal precision requires $\sim10^{32}$ bits. As of 2021, the world's
fastest classical supercomputer \citep{monroe2020fugaku}, Fugaku,
has an enormous $10^{16}$ bit memory. However, this is a factor of
$10^{16}$ too small even to store the quantum state, let alone compute
high order correlations. Such limitations prevent the direct use of
exact number state methods for calculations in large quantum networks.

In one-dimensional lattices with low excitations and nearest-neighbor
couplings, one may approximate a large network by many smaller ones.
This gives rise to approximations known as the density matrix renormalization
group, or tensor network methods \citep{Schollwock_RMP2005_Review_DMRG}.
Such methods can fail when there is increased connectivity. An example
of this is if there is increased dimensionality, or extended correlations
across the entire network, which may lead to all of the modes being
correlated. 

Fortunately, there are other techniques available. One can carry out
probabilistic simulations in a phase-space which allows a complete
representation of the density matrix. Wigner \citep{Wigner1932Quantum}
first showed how to map quantum wave-functions onto a distribution
function on a phase-space of classically meaningful quantities, although
this is only probabilistic in some cases. The general concept of a
phase-space representation in quantum mechanics is a correspondence
between a set of phase-space variables and a set of quantum operators
with a particular ordering \citep{Dirac_RevModPhys_1945}. Subsequent
work was motivated by the idea \citep{Moyal_1949} that one can use
such methods to calculate the dynamics of the observables. It is these
that are measurable and of most scientific interest. 

The first probabilistic quantum phase-space distribution was developed
by Husimi \citep{Husimi1940}, using the concept of a coherent state
\citep{Schrodinger_CS,Glauber_1963_P-Rep}. Later this was shown to
lead to an anti-normally ordered bosonic field representation \citep{Cahill_Glauber_1969_Density_operators},
called the Q-function, which provides a positive representation of
quantum fields. Normally ordered representations on a classical phase-space,
called P-representations, are useful for simulating lasers, but are
singular for non-classical states. There is a range of classical phase-space
methods of this type, often called $s$-ordered \citep{Cahill_Glauber_1969_Density_operators}.
Glauber's P-representation was later extended to a non-classical phase-space
with double the classical dimensions, called the positive P-representation.
This is normally ordered and always exists as a smooth, positive distribution
for any quantum state \citep{Drummond_generalizedP1980,Drummond:2016}.

The positive P-representation gives results with comparable sampling
errors to photon-counting experiments. In both cases, one measurement
or one random sample must be repeated many times to give overall averages.
There are statistical errors, but experimental measurements have the
same issue. Other phase-space representations can be singular or negative,
or have larger sampling errors for photon counting. Wigner phase-space
representations are best for a special case: Gaussian initial states
combined with linear or weakly nonlinear networks and homodyne detectors,
rather than photon-counting \citep{Drummond:1993EPL}.

There are also proposals for similar techniques using modified Wigner
functions in qubit-based systems \citep{mari2012positive,veitch2014resource}
which are largely outside the scope of this tutorial, and are not
treated here. 

\subsection{Wigner representation}

The oldest method for quantum phase-space is the Wigner representation
of bosonic states \citep{Wigner1932Quantum,Cahill_PhysRev1969}. This
provides a method for random simulations, provided the Wigner distribution
is positive and has a corresponding stochastic process. Such techniques
can treat linear networks or nonlinear quantum networks at large occupation
numbers \citep{Drummond:1993EPL,corney2008simulations}, in excellent
agreement with experiment. 

This distribution can be written as:
\begin{equation}
W\left(\boldsymbol{\alpha}\right)=\frac{1}{\pi^{2N}}\int d^{2}\mathbf{z}\left\langle e^{i\mathbf{z}\cdot\left(\hat{\mathbf{a}}-\boldsymbol{\alpha}\right)+i\mathbf{z}^{*}\cdot\left(\hat{\mathbf{a}}^{\dagger}-\boldsymbol{\alpha}^{*}\right)}\right\rangle .
\end{equation}
Operator mean values are obtained using a mapping with symmetric ordering,
of form:
\begin{align}
\left\langle \hat{a}_{i}^{\dagger m}\dots\hat{a}_{j}^{n}\right\rangle _{SYM} & =\int d^{2}\boldsymbol{\alpha}\left[\alpha_{i}^{*m}\ldots\alpha_{j}^{n}\right]W\left(\boldsymbol{\alpha}\right).
\end{align}

This is a particularly useful approach for networks that employ homodyne
detectors. These measure quadratures $\hat{x}_{i}^{\theta_{i}}=\hat{a}e^{-i\theta_{i}}+\hat{a}^{\dagger}e^{i\theta_{i}}$,
which are commonly used to characterize squeezing or entanglement. 

For a positive Wigner function, the symmetrically ordered moments
of quadrature readouts are equivalent to random samples of the Wigner
distribution \citep{Cahill_Glauber_1969_Density_operators}, with
\begin{align}
\left\langle \hat{x}_{i}^{\theta_{i}}\dots\hat{x}_{j}^{\theta_{j}}\right\rangle _{SYM} & =\left\langle x_{i}^{\theta_{i}}\ldots x_{j}^{\theta_{j}}\right\rangle _{W}.
\end{align}
Here, the $x$ and $p$ quadratures are defined as
\begin{align}
\hat{x}_{i}=\hat{x}_{i}^{0} & =\hat{a}+\hat{a}^{\dagger},\nonumber \\
\hat{p}_{i}=\hat{x}_{i}^{\pi/2} & =i\left(\hat{a}^{\dagger}-\hat{a}\right).
\end{align}

Provided the initial state is Gaussian and either the Hamiltonian
is linear or the photon numbers are large, the Wigner distribution
is positive and can be readily sampled stochastically. The calculation
of the moments is efficient, scaling linearly with the system-size,
provided the samples are available. A Wigner simulation under these
conditions is directly equivalent to an experimental readout of measured
quadratures \citep{Drummond:1993EPL,bartlett2002efficient}, where
$x_{i}^{\theta_{i}}=\alpha e^{-i\theta_{i}}+\alpha^{*}e^{i\theta_{i}}$
corresponds to the measured quadrature current. The reason for this
is that third-order terms in the propagation equations vanish either
in the limit of linear couplings or large photon number, so an initially
positive distribution, such as a Gaussian distribution, will remain
positive. To make use of this property one needs to generate random
samples, which is treated below for the case of linear network experiments.

These conditions are very restrictive, however. Linear network experiments
can also use particle detectors, which are equivalent to normal ordering.
Such detectors can't be efficiently treated using the Wigner representation,
unless corrections are included which transform the ordering from
symmetric to normal. However, the additional vacuum noise inherent
in the Wigner representation means that there is a rapidly growing
sampling error for high order correlations, reducing the scalability.
Similarly, if there are nonlinearities, a Wigner simulation will omit
certain nonlinear quantum noise corrections. 

The other phase-space representations such as the Husimi Q-function,
for anti-normal ordering, and Glauber's P-representation, for normal
ordering, have their own drawbacks. Similar to the Wigner function,
the Q-function suffers from large sampling errors while the P-representation
has highly singular distributions for the non-classical inputs present
in linear network experiments. 

\subsection{Generalized P-representation}

The generalized P-representation \citep{Drummond_generalizedP1980,opanchuk2019robustness}
was developed as the quantum successor to Glauber's P-representation,
as it always has a positive distribution for any quantum state. This
is the most efficient representation for simulating normally-ordered
photon detectors, due to its ability to use efficient stochastic sampling
methods to generate input probabilities.  

This method expands the density matrix over a multi-dimensional subspace
of the complex plane, defined as:
\begin{equation}
\hat{\rho}=\Re\iint P\left(\boldsymbol{\alpha},\boldsymbol{\beta}\right)\hat{\Lambda}\left(\boldsymbol{\alpha},\boldsymbol{\beta}\right)\mathrm{d}\mu\left(\boldsymbol{\alpha},\boldsymbol{\beta}\right),\label{eq:generalised P}
\end{equation}
where, by only taking the real part, the output is always hermitian.
The basis operator 
\begin{equation}
\hat{\Lambda}\left(\boldsymbol{\alpha},\boldsymbol{\beta}\right)=\frac{|\boldsymbol{\alpha}\rangle\langle\boldsymbol{\beta}^{*}|}{\langle\boldsymbol{\beta}^{*}|\boldsymbol{\alpha}\rangle},
\end{equation}
projects the density matrix onto multi-mode coherent states \citep{Glauber1963_CoherentStates},
while $\mathrm{d}\mu\left(\boldsymbol{\alpha},\boldsymbol{\beta}\right)$
is an integral measure on the $2M$-dimensional complex space of coherent
state amplitudes $\boldsymbol{\alpha},\boldsymbol{\beta}$. The definition
of the integral measure determines whether the distribution $P\left(\boldsymbol{\alpha},\boldsymbol{\beta}\right)$
is positive or complex valued. We use the positive distribution here
where $\mathrm{d}\mu\left(\boldsymbol{\alpha},\boldsymbol{\beta}\right)\equiv d^{2}\boldsymbol{\alpha}d^{2}\boldsymbol{\beta}$
corresponds to a $4M$-dimensional real Euclidean volume. 

Number state inputs can also be simulated, but for greater sampling
efficiency, the complex P-representation is preferable for such inputs.
To simulate these cases, an additional complex factor $\Omega$ should
be used, when sampling the distribution \citep{opanchuk2018simulating,drummond2020initial}.

Besides computational efficiency, the generalized P-representation
has the added benefit of making normally ordered operator moments
equal to stochastic moments \citep{Drummond_generalizedP1980,gardiner2004handbook,drummond2014quantum}:
\begin{equation}
\left\langle \hat{a}_{j_{1}}^{\dagger},\dots,\hat{a}_{j_{n}}\right\rangle =\left\langle \beta_{j_{1}},\dots,\alpha_{j_{n}}\right\rangle _{P},
\end{equation}
where quantum expectation values are denoted $\left\langle \right\rangle $,
and averages obtained using the generalized P-representation are $\left\langle \right\rangle _{P}$.
This allows one to simply replace quantum operators with stochastic
variables without changing the analytical solutions. As in the Wigner
case, high-order moments are efficiently calculable in times that
scale linearly with system-size, provided samples are available.

The following sections demonstrate how to efficiently generate the
samples, which have now reached $16,000$ modes. The main memory limitation
is the storage of the experimental $M\times M$ transmission matrix
data, explained below.

\subsection{Simulating Gaussian inputs}

Any algorithm that simulates correlations of a complex system such
as a linear bosonic network must be tested against exactly known distributions.
Provided they have positive distributions, the Wigner and generalized
P-representation are both well suited to simulating Gaussian distributions
such as thermal and squeezed distributions. The phase-space distributions
are themselves Gaussian in these cases. 

To simulate any phase-space representation, one must first generate
stochastic samples. For Gaussian quantum inputs into a network, the
corresponding phase-space representations are also Gaussian. As a
result, they are completely defined by their means - which we take
as zero here - and their second-order correlations. 

We define an ordering parameter $\sigma$ to signify the amount of
vacuum noise in an operator ordering. Compared to the $s$-ordering
in earlier conventions \citep{Cahill_PhysRev1969}, $\sigma=\left(1-s\right)/2$.
Thus, $\sigma=0$ signifies normal ordering, $\sigma=\frac{1}{2}$
symmetric ordering, and $\sigma=1$ anti-normal ordering. For independent
quantum inputs, with no correlations between the real and imaginary
quadratures, we define the general $\sigma$-ordered quadrature variance
as 
\begin{align}
\left\langle \left\{ \left(\Delta\hat{x}_{j}\right)^{2}\right\} _{\sigma}\right\rangle  & =\left(\Delta_{\sigma j}^{x}\right)^{2}=2\left(n_{j}+\sigma+\tilde{m}_{j}\right)\nonumber \\
\left\langle \left\{ \left(\Delta\hat{y}_{j}\right)^{2}\right\} _{\sigma}\right\rangle  & =\left(\Delta_{\sigma j}^{y}\right)^{2}=2\left(n_{j}+\sigma-\tilde{m}_{j}\right).\label{eq:sigma-ordered samples}
\end{align}
Here, $n_{j}=\left\langle \hat{a}_{j}^{\dagger}\hat{a}_{j}\right\rangle $
and $\tilde{m}_{j}=\left\langle \hat{a}_{j}\hat{a}_{j}\right\rangle $
which are defined in terms of linear bosonic networks in Section III
A. Since the phase-space quadratures are:
\begin{equation}
x_{i}^{\theta_{i}}=\alpha e^{-i\theta_{i}}+\beta e^{i\theta_{i}},
\end{equation}
we can now generate stochastic samples in any representation by simply
matching the $\sigma$ parameter to the type of representation and
generating random Gaussian samples from
\begin{align}
\alpha_{j} & =\left(\Delta_{\sigma}^{x}w_{j}+i\Delta_{\sigma}^{y}w_{j+M}\right)/2\nonumber \\
\beta_{j} & =\left(\Delta_{\sigma}^{x}w_{j}-i\Delta_{\sigma}^{y}w_{j+M}\right)/2.\label{eq:input-sampling-Gaussian}
\end{align}
The terms $w_{j}$ are uncorrelated Gaussian real random numbers with
unit variance, thus requiring $2M$ random Gaussian numbers per phase-space
sample. This is the most computationally efficient method for generating
samples of each representation. Here $\beta\equiv\alpha^{*}$ in the
case of the Wigner representation, which uses a classical phase-space,
with $\sigma=1/2$. The positive P-representation has $\sigma=0$,
and can have $\beta\neq\alpha^{*}$ if either of $\Delta_{\sigma}^{x}\,$
or $\,\Delta_{\sigma}^{y}$ from Eq.(\ref{eq:sigma-ordered samples})
are imaginary. Here $\Delta_{\sigma}^{x}\,$ or $\,\Delta_{\sigma}^{y}$
can be imaginary if $n_{j}<\left|\tilde{m}_{j}\right|$, in the positive
P-representation. 

These variances are related to the squeezed-state parameters in the
next section. Phase-space simulations do not always replicate observable
random outputs. This depends on the type of mapping used. In this
tutorial we will first treat normally ordered mappings for photon
counting detectors, which replaces an experimental discrete integer
variable by a complex variable with identical correlations. We will
also consider Wigner representations, where the stochastic process
is indistinguishable from an experimental output. 

When conducting simulations in phase-space one must obtain averages
over a large number of samples, and for any ensemble average to be
reliable, the calculation must be repeated multiple times for each
input mode. Generally, the more the simulation is repeated, the more
accurate the ensemble averaged output will be, with smaller errors. 

The sampling error $e_{s}$ scales with the number of samples as $1/\sqrt{S}$
\citep{drummond2020initial}. For any accurate simulation of an experimental
system, the theoretical sampling error must be smaller than the experimental
sampling error. One can either continuously increase the number of
samples computed in parallel, or else one can simply increase the
number of times the calculation is repeated, although this does increase
computation time significantly, if multi-core computations aren't
possible.

\section{Linear bosonic networks}

Linear bosonic networks may appear trivial, since the quantum correlations
have exact analytic solutions. However, when mode numbers are large,
these analytic solutions can become exponentially hard to compute,
resulting in matrix permanents \citep{Aaronson2011}, Hafnians \citep{Hamilton2017PhysRevLett.119.170501}
or Torontonians \citep{quesada2018gaussian}. Consequently, most output
correlations cannot be exactly computed in large systems with $M\apprge60$
modes. 

In recent experiments with $M=100$ modes \citep{zhong2020quantum},
a direct simulation was estimated, in 2019, to take nearly a billion
years, with the world's fastest supercomputer. Improved algorithms
have now been proposed to reduce the classical simulation time \citep{bulmer2021boundary}.
These are also exponentially slow, meaning that they become impractical
with even small size increases. 

The cause of the problem is that the analytic expressions for correlations
may involve combinatoric expressions with exponentially many terms.
Therefore, having an analytic solution does not eliminate the problem
of Hilbert space dimension. These solutions can still take exponentially
long to calculate. Just as seriously, they produce round-off errors
which grow rapidly \citep{wu2018benchmark}. This is the basis for
the boson sampling quantum computer. These devices generate random
numbers that cannot be generated efficiently by classical means \citep{Aaronson2011,AaronsonArkhipov2013LV},
by measuring photon detection events, or counts, in each mode.

Since one cannot compute the output correlations for large networks
exactly, an intriguing issue arises. How does one verify that the
computer is working? This is similar to testing problems related to
environmental standards \citep{boretti2017future}. One has to ask:
was the test carried out correctly? With quantum computers there is
a new question: can a test be carried out at all? Many proposed tests
are restricted to low order correlations \citep{Walschaers2016_NJP18,phillips2019benchmarking,huang2017statistical}
which cannot uniquely characterize the output distributions, because
these in general require all orders to characterize \citep{Moran1984}.
Other proposed tests require different experimental configurations,
meaning that the actual GBS data is not used \citep{chabaud2021efficient}. 

This can be answered, at least in part, through the use of sampled
correlations and probabilities, as explained in the next section.

\begin{figure}
\begin{centering}
\includegraphics[width=0.9\columnwidth]{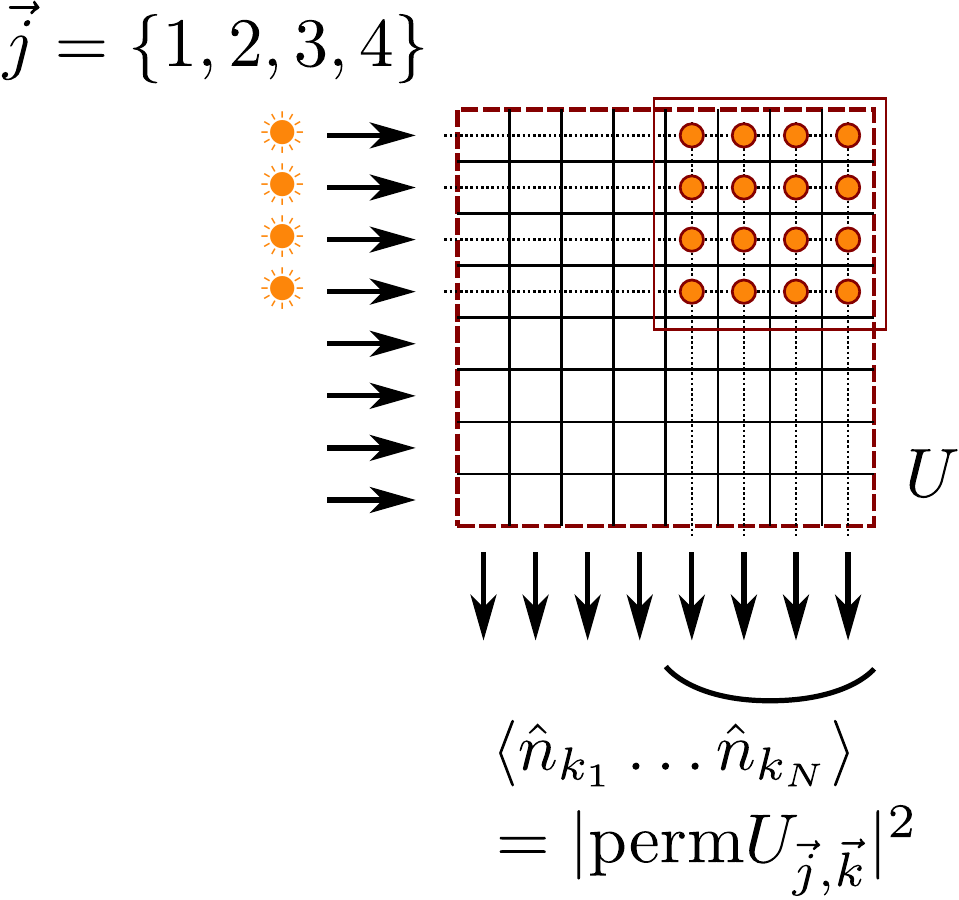}
\par\end{centering}
\caption{Schematic of a boson sampling experiment. This shows a simple, $8\times8$
unitary transformation with four single photon inputs. A measurement
of the correlated outputs is given by the permanent function which
is exponentially hard to compute exactly for large matrices. The much
larger, more recent GBS experiments use squeezed inputs, have decoherence,
and employ highly efficient on-off detectors. \label{fig:boson-sampling-diagram}}
\end{figure}

\subsection{Input quantum states}

The original iteration of the boson sampling quantum computer sent
single photon Fock states into an optical network, the output of which
is then measured using photon detectors \citep{AaronsonArkhipov:2011}.
An inability to scale the size of the network to computationally interesting
regimes due to constraints in generating single, indistinguishable
photons \citep{broome2013photonic,spring2013boson,Wang2019BosonSampling}
has driven a push to seek alternative, experimentally accessible bosonic
networks. 

One such network replaces the number state with a Gaussian single-mode
squeezed state as the nonclassical light source \citep{Hamilton2017PhysRevLett.119.170501}.
This type of linear bosonic network is called a Gaussian boson sampler
(GBS). It has recently been implemented in multiple large scale experiments
with $100$ modes or more, claiming to calculate samples with a Torontonian
distribution \citep{zhong2020quantum}. A GBS sends $N\subset M$
single-mode squeezed states into an optical network which is described
by the linear, unitary transformation matrix $\boldsymbol{T}$. The
input modes are transformed by the matrix into $M$ output modes,
allowing measurements to be made on the output state $\hat{\rho}^{(out)}$. 

The input state $\hat{\rho}^{(in)}$ can be defined as a product of
single-mode pure squeezed states 
\begin{equation}
\hat{\rho}^{(in)}=\prod_{j}\left|\boldsymbol{r}_{j}\right\rangle \left\langle \boldsymbol{r}_{j}\right|,\label{eq:input density matrix}
\end{equation}
with squeezing vector $\boldsymbol{r}=\left[r_{1},\dots,r_{N}\right]$,
if each input mode $N$ is independent. Using the squeezed state expansion
$\left|r\right\rangle =\hat{S}\left(r\right)\left|0\right\rangle $,
where $\hat{S}\left(r\right)$ is the squeezing operator, the input
photon number can be found from the expectation of the single-mode
number operator 
\begin{align}
n\left(r_{j}\right) & =\left\langle \hat{a}_{j}^{\dagger}\hat{a}_{j}\right\rangle \nonumber \\
 & =\sinh^{2}\left(r_{j}\right).
\end{align}
The corresponding coherence of the input photons can similarly be
defined as 
\begin{align}
m\left(r_{j}\right) & =\left\langle \hat{a}_{j}^{2}\right\rangle \nonumber \\
 & =\sinh\left(r_{j}\right)\cosh\left(r_{j}\right).
\end{align}

The normally ordered variances of the quadrature operators $\hat{x}_{j}=\hat{a}_{j}+\hat{a}_{j}^{\dagger}$
and $\hat{y}_{j}=\left(\hat{a}_{j}-\hat{a}_{j}^{\dagger}\right)/i$,
with commutation relation $\left[\hat{x}_{l},\hat{y}_{j}\right]=2i\delta_{lj}$,
are well documented for pure squeezed states and can be written in
terms of the input photon number and coherence for the generalized
P-distribution, as 

\begin{align}
\left\langle \left(\Delta\hat{x}_{j}\right)^{2}\right\rangle  & =2\left(n+m\right)=e^{2r_{j}}\nonumber \\
\left\langle \left(\Delta\hat{y}_{j}\right)^{2}\right\rangle  & =2\left(n-m\right)=e^{-2r_{j}}.\label{eq:pure state variances}
\end{align}
This is easily expressed in any ordering as 

\begin{align}
{\normalcolor \left\langle \left\{ \left(\Delta\hat{x}_{j}\right)^{2}\right\} _{s}\right\rangle } & {\normalcolor =2\left(n+\sigma+m\right)}\nonumber \\
{\normalcolor \left\langle \left\{ \left(\Delta\hat{y}_{j}\right)^{2}\right\} _{s}\right\rangle } & {\normalcolor =2\left(n+\sigma-m\right).}\label{eq:s-ordered variance, no decoherence}
\end{align}

Experiments that generate squeezed states usually have additional
decoherence caused by longitudinal mode mismatching \citep{zhong2020quantum}
or  other dephasing effects \citep{drummond2020initial}. To accurately
simulate bosonic networks one must include decoherence. An approximate
model for decoherence consists of an intensity transmissivity $T=1-\epsilon$,
which reduces the input photon intensity while adding $n_{j}^{th}=\epsilon n_{j}$
thermal photons per mode. Although the output photon number is unchanged,
the coherence of each mode is now given by 
\begin{equation}
\left\langle \hat{a}_{j}^{2}\right\rangle =\tilde{m}=\left(1-\epsilon\right)m\left(r_{j}\right).
\end{equation}

The addition of decoherence alters the input modes, which can no longer
be considered simply pure squeezed states, and have variances given
by Eq (\ref{eq:sigma-ordered samples}), which now include decoherence
in each mode. Although the relationship is similar to Eq (\ref{eq:pure state variances}),
the variances can no longer be simplified to $\exp\left(\pm2r_{j}\right)$,
which increases their product above the Heisenberg limit. These more
realistic states are also called thermal squeezed states in the literature,
usually with other parameterizations \citep{Fearn_JModOpt1988}.

\subsection{Sampling squeezed phase-space distributions}

The normal ordering of the photo-detectors used to measure the output
state, and its non-singular nature for squeezed states makes the generalized
P-representation the most efficient phase-space representation to
simulate GBS. However, before we can sample the output state, we must
first generate the input state in the generalized P-representation. 

This is straightforwardly applied by initially neglecting the added
input mode decoherence, this is added later with the beam splitter
model of decoherence described above when the stochastic samples are
generated. Since our input is a pure squeezed state, and recalling
that a squeezed state corresponds to a superposition of coherent states,
we can expand the density matrix, Eq.(\ref{eq:input density matrix}),
using the coherent state expansion \citep{janszky1990squeezing}

\begin{equation}
\left|r\right\rangle =C\int_{-\infty}^{\infty}\mathrm{d}\alpha_{x}\exp\left(-\alpha_{x}^{2}/\gamma\right)\left|\alpha_{x}\right\rangle .
\end{equation}
Here, $\left|\alpha_{x}\right\rangle $ is a Glauber coherent state
\citep{Glauber_1963_P-Rep}, $\alpha_{x}$ in this case is a real
number and $C=(1+\gamma)^{1/4}/\sqrt{\pi\gamma}$ is the normalization
constant with $\gamma=e^{2r}-1$. 

Once expanded, the input density matrix is given by:
\begin{equation}
\hat{\rho}^{(in)}=\Re\iint P\left(\boldsymbol{\alpha}_{x},\boldsymbol{\beta}_{x}\right)\hat{\Lambda}\left(\boldsymbol{\alpha}_{x},\boldsymbol{\beta}_{x}\right)\mathrm{d}\boldsymbol{\alpha}_{x}\mathrm{d}\boldsymbol{\beta}_{x},\label{eq:input state P-rep}
\end{equation}
with corresponding quadrature variables,
\begin{align}
\boldsymbol{x} & =\boldsymbol{\alpha}+\boldsymbol{\beta}\nonumber \\
\boldsymbol{p} & =\frac{\left(\boldsymbol{\alpha}-\boldsymbol{\beta}\right)}{i}.
\end{align}
and an input distribution:
\begin{equation}
P\left(\boldsymbol{\alpha},\boldsymbol{\beta}\right)=\prod_{j}C_{j}e^{-\left(\alpha_{xj}^{2}+\beta_{xj}^{2}\right)\left(\gamma_{j}^{-1}+1/2\right)+\alpha_{xj}\beta_{xj}}\delta\left(\alpha_{yj}\right)\delta\left(\beta_{yj}\right),\label{eq:input probability}
\end{equation}
where $C_{j}=\sqrt{1+\gamma_{j}}/\left(\pi\gamma_{j}\right)$ is the
normalization constant. The distribution $P\left(\boldsymbol{\alpha},\boldsymbol{\beta}\right)$
is a positive, Gaussian distribution which is restricted to the real
axes and is obtained using the inner product of two coherent states
\begin{equation}
\left\langle \beta_{x}\mid\alpha_{x}\right\rangle =e^{\alpha_{x}\beta_{x}-\frac{1}{2}\left(\alpha_{x}^{2}+\beta_{x}^{2}\right)},
\end{equation}
which is also needed to construct the projection operator. For any
input defined as a product of single-mode squeezed states, Eq.(\ref{eq:input probability})
always exists. 

In order to simulate any phase-space representation, one must first
generate stochastic samples of the initial quantum state. However,
each phase-space representation samples the initial state differently,
leading to different sampling errors for the same initial state depending
on the representation. Unlike the Wigner and Q-function representations,
the generalized P-representation has a much smaller sampling error
due to it not containing additional vacuum noise \citep{drummond2020initial}.
This is another reason why the generalized P-representation is the
most efficient representation to simulate GBS. 

Generating stochastic samples for the pure-state input distribution,
Eq. (\ref{eq:input probability}), is straightforward for Gaussian
states. Using the real Gaussian noises $\left\langle w_{i}w_{j}\right\rangle _{P}=\delta_{ij}$,
we wish to construct the random samples $\overrightarrow{\alpha}=\left[\boldsymbol{\alpha},\boldsymbol{\beta}\right]=\left[\alpha_{1},\dots,\alpha_{2M}\right]$
which is achieved using the stochastic model given above for a pure
squeezed state \citep{drummond2020initial} 
\begin{align}
\alpha_{j}^{0} & =\delta_{j+}w_{j}+\delta_{j-}w_{j+M}\nonumber \\
\beta_{j}^{0} & =\delta_{j+}w_{j+M}+\delta_{j-}w_{j}.
\end{align}
Here, $\delta_{j\pm}$ is a real Gaussian sample with solution $\delta_{j\pm}^{2}=\frac{1}{2}\left(m_{j}\pm\sqrt{n_{j}}\right)$.
However, since our input state isn't a simple pure squeezed state,
we must alter the stochastic model to include the intensity transmissivity
and additional thermal photons per mode. The full stochastic random
samples are therefore expressed as in the general form of Eq (\ref{eq:input-sampling-Gaussian}).
The additional decoherence has the added effect of changing our input
distribution, which is no longer always restricted to the real axes
in thermalized cases. 

\subsection{Effect of the linear network}

Once the input stochastic samples have been generated, they are transformed
using the linear transformation matrix $\boldsymbol{T}$. We define
this as an $M\times M$ complex matrix, which is measured in the experimental
network. The transformed amplitudes, $\boldsymbol{\alpha}'=\boldsymbol{T}\boldsymbol{\alpha}$
and $\boldsymbol{\beta}'=\boldsymbol{T}^{*}\boldsymbol{\beta}$, are
then used to sample the output state 
\begin{equation}
\hat{\rho}^{(out)}=\Re\iint P\left(\boldsymbol{\alpha},\boldsymbol{\beta}\right)\hat{\Lambda}\left(\boldsymbol{T}\boldsymbol{\alpha},\boldsymbol{T}^{*}\boldsymbol{\beta}\right)\mathrm{d}\mu\left(\boldsymbol{\alpha},\boldsymbol{\beta}\right),\label{eq:output state P-rep}
\end{equation}
from which photon counting measurements are taken. For a lossless
network, $\boldsymbol{T}$ is a unitary matrix, but generally it is
non-unitary due to losses. Since coherent states remain coherent even
with absorption, the above transformation is still valid for the positive
P-representation, even when $\boldsymbol{T}$ is not unitary. For
the Wigner representation, one can transform the amplitudes so that
$\boldsymbol{\alpha}'=\boldsymbol{T}\boldsymbol{\alpha}$ only in
the unitary case; otherwise additional noise terms are needed \citep{drummond2021simulating}. 

\section{Measurements in Gaussian boson sampling}

The photo-detectors, or click detectors, used in boson sampling experiments
saturate for more than one count at a detector. To clarify, a ``click''
is a detection event, while a ``count'' is the number of measured
photons. Each detector therefore outputs binary numbers $c_{j}=0,1$
which, for $M$ modes, produces point patterns represented as a vector
$\boldsymbol{c}$. There are $2^{M}$ possible patterns available,
where total probability of any individual click pattern is therefore
$2^{-M}$, which is exponentially small for large $M$, provided that
they are equally probable.  

Each detector is defined by the projection operator 
\begin{equation}
\hat{\pi}_{j}\left(c_{j}\right)=:e^{-\hat{n}'_{j}}\left(e^{-\hat{n}'_{j}}-1\right)^{c_{j}}\colon,\label{eq:single detector projection operator}
\end{equation}
the expectation of which, $\left\langle \hat{\pi}_{j}\left(c_{j}\right)\right\rangle $,
is the probability of observing a single count at the $j$-the detector,
where $\hat{n}'_{j}$ is the output photon number. For a bosonic network
with $M$ detectors \citep{Sperling2012True}, the multi-click projection
operator becomes:
\begin{equation}
\hat{\Pi}\left(\boldsymbol{c}\right)=\bigotimes_{j=1}^{M}\hat{\pi}_{j}\left(c_{j}\right),\label{eq:Torontonian}
\end{equation}
which is the product of projection operators at each detector. For
Gaussian inputs, the expectation, $\left\langle \hat{\Pi}\left(\boldsymbol{c}\right)\right\rangle $,
is the probability of observing any single binary pattern. This is
the Torontonian function \citep{quesada2018gaussian}. For example,
if $M=6$, the binary pattern $\left[1,1,1,0,0,0\right]$ has the
projector: 

\begin{equation}
\hat{\Pi}\left(\boldsymbol{c}\right)=\hat{\pi}_{1}\left(1\right)\hat{\pi}_{2}\left(1\right)\hat{\pi}_{3}\left(1\right)\hat{\pi}_{4}\left(0\right)\hat{\pi}_{5}\left(0\right)\hat{\pi}_{6}\left(0\right).
\end{equation}
which is a product of three click projectors in the first three detectors
followed by three non-click projections. This binary pattern has a
probability of $2^{-6}\approx0.016$ which, although small, is still
easily computed classically. 

Upon inspection, it is clear how Eq. (\ref{eq:Torontonian}) becomes
exponentially complex as $M$ increases, from recent experiments \citep{zhong2020quantum}
an $M=100$ network has a probability of $2^{-100}\approx10^{-31}$
of observing a single binary pattern. One is more likely to win the
lottery multiple times than to observe a single binary pattern. We
have now arrived at the crux of the verification problem with linear
bosonic networks. How can the outputs be verified when the probability
of obtaining any individual click pattern is nearly zero? We note
that it is known that there are computational reasons why one cannot
directly classically generate such counts in a reasonable time. However,
the sparseness of the probabilities is still be an issue even if one
could classically generate them.

To answer this, we use binning. Binning is the common approach of
combining large, sparse sums of data into discrete bins in order to
obtain reproducible statistics. Applying it to the output of a bosonic
network, one can combine exponentially many click patterns into bins
based on the number of photon counts. This allows one to calculate
grouped count probabilities, which is the probability of observing
$m$ counts in any pattern, as opposed to explicitly simulating discrete
experimental photon counts. 

Consider the probability of observing $\boldsymbol{m}=\left(m_{1},\dots,m_{g}\right)$
grouped counts in $S_{1},S_{2},\dots$ disjoint sets of $\boldsymbol{M}=\left(M_{1},M_{2},\dots\right)$
output modes, while disregarding the counts in any other modes. This
is a marginal probability if $n=\sum_{j=1}^{g}M_{j}<M$, and it requires
summing over all patterns which satisfy the condition that the counts
for the detectors in the set $S_{1}$ sum up to $m_{1}$, the counts
in $S_{2}$ sum up to $m_{2}$, or in general, that:
\begin{equation}
m_{j}=\sum_{i\in S_{j}}c_{i}.
\end{equation}

This result is achieved on defining a general grouped count correlation
as 
\begin{equation}
\mathcal{G}_{M}^{(n)}\left(\boldsymbol{m}\right)=\left\langle \prod_{j=1}^{g}\left[\sum_{\sum c_{i}=m_{j}}\hat{\Pi}_{S_{j}}\left(\boldsymbol{c}\right)\right]\right\rangle .\label{eq:grouped count probability}
\end{equation}
 Here, $n=\sum_{j=1}^{g}M_{j}\leq M$ is the total correlation order
following the standard definition of Glauber \citep{Glauber1963_CoherentStates}.
For example, if there is only one mode, it is a first order correlation,
if there are two modes it is a second-order correlation. The full
probability of a given pattern, the Torontonian, is obtained by taking
$n=M$ and $\boldsymbol{M}=\left(1,1,\dots\right)$. 

Applications to various experimental correlations are easily illustrated.
This approach is essential for experiments. By grouping patterns together,
one can transform a sparse distribution to a non-sparse one, for which
probabilities can be estimated from the data. For example, the probability
of observing a single count at the $j$-th detector, is obtained when
$n=1$ and $M_{j}=1$, giving $\mathcal{G}_{1}^{(1)}\left(\boldsymbol{m}\right)=\left\langle \hat{\pi}_{j}\left(c_{j}\right)\right\rangle $.
This corresponds to the expectation of Eq. (\ref{eq:single detector projection operator}).
Similarly, one can sum over all $n-th$ order correlations.

At first glance, it might appear that grouped correlations only introduce
added difficulty to the problem of verification. If one cannot compute
the Torontonian, how is it possible to sum over exponentially many
different Torontonians? This seems just like asking a runner who has
never finished a marathon, to run one every day for a year. However,
with phase-space methods the computational problem is sampling error,
rather than the order itself. By adding many terms, the relative sampling
error is reduced. This makes this hard calculation feasible. 

To show how this works, note that since $\hat{\pi}_{j}$ is normally
ordered, it can be readily simulated by simply replacing it with the
phase-space observable 
\[
\pi_{j}\left(c_{j}\right)=e^{-n'_{j}}\left(e^{-n'_{j}}-1\right)^{c_{j}},
\]
where $n'_{j}$ is sampled from the output state, Eq.(\ref{eq:output state P-rep}),
with probability $P\left(\boldsymbol{\alpha},\boldsymbol{\beta}\right)$.
However the question still remains about how to efficiently calculate
an exponentially large summation. Fortunately, this is achievable,
through the use of Fourier transforms, as shown in earlier work on
number-counting methods for boson sampling \citep{opanchuk2018simulating}.
By introducing the angles $\theta_{i}=2\pi/\left(M_{i}+1\right)$
together with the Fourier observable 
\begin{equation}
\tilde{\mathcal{G}}_{M}^{(n)}\left(\boldsymbol{k}\right)=\left\langle \prod_{j=1}^{g}\bigotimes_{i\in S_{j}}\left(\pi_{i}\left(0\right)+\pi_{i}\left(1\right)e^{-ik_{i}\theta_{i}}\right)\right\rangle _{P},
\end{equation}
with $k_{i}=0,\dots m_{i}$, one can use a multi-dimensional, inverse
discrete Fourier transform to calculate the grouped correlations as
\begin{equation}
\mathcal{G}_{M}^{(n)}\left(\boldsymbol{m}\right)=\frac{1}{\prod\left(M_{i}+1\right)}\sum_{\boldsymbol{k}}\tilde{\mathcal{G}}_{M}^{(n)}\left(\boldsymbol{k}\right)e^{i\sum k_{i}\theta_{i}m_{i}}.\label{eq:Fourier grouped correlation}
\end{equation}
The inverse Fourier transform reduces the computational complexity
of the sums by removing all patterns which don't contain $\boldsymbol{m}$
counts. Therefore, grouped correlations can be used obtain measurable,
statistical fingerprints of an experimental network that are also
computable. 

\subsection{Phase-space algorithm}

So far we have described the theoretical tools necessary to verify
the output correlations of a GBS network using grouped correlations
sampled from the generalized P-representation. An efficient algorithm
has been developed which implements these tools and was recently applied
to a $100$-mode GBS experiment by Zhong et al \citep{zhong2020quantum},
showing excellent agreement between theory and experiment when decoherence
was included \citep{drummond2021simulating}. In this section we will
outline the details of this algorithm and present results of various
tests that were conducted to verify the simulated correlations.

Simulating ensemble averaged grouped correlations using the discrete
Fourier transform technique introduced in Eq. (\ref{eq:Fourier grouped correlation})
is straight forward and computationally efficient for most observables.
One must simply loop over the count projector $\pi_{j}$, taking the
product over input samples before calculating the ensemble average
for each $k_{i}$. The discrete Fourier transform is then applied
to all ensemble averages giving the grouped correlations of a particular
observable. 

Multi-dimensional binning of the grouped correlations is more challenging,
although also giving far more points of comparison. This is highly
desirable, to provide a means to distinguish experimental data from
direct classical simulations with 'fake' algorithms. A grouped correlation
can be binned into $D$-dimensions, giving the probability of observing
$n=m_{1},\dots,m_{M/D}$ counts in each dimension. To simulate this,
the input samples must be partitioned into bins corresponding to the
number of dimensions. The count projector is then calculated for each
$k_{i}$ as before, however this occurs over each partitioned separately
before a multi-dimensional discrete Fourier transform is applied to
the ensemble average of each partition. 

For smaller dimensions, this does not take a significant amount of
time to compute. However, as the dimensionality increases, up to $M$
dimensions being theoretically possible, the computation time increases
significantly. Memory constraints from the increased ensemble size
start to occur. Another practical issue is that increased dimensionality
leads to fewer experimental or computational samples per bin. If either
of these are too small, the comparisons lose significance. This requires
an understanding of statistical tests.

When verifying correlations of large complex experimental networks,
one must make sure sufficient statistical tests are conducted to verify
the simulated correlations. Although this is the case with any comparison
of theory with experiment, it is even more important when decoherence
is included, as small alterations in the input mode decoherence can
greatly affect the output grouped correlations. To quantify this,
a chi-square test is used to compare phase-space simulations with
GBS experimental correlations. Chi-square tests are commonly used
to calculate the differences between theoretical and experimental
measurements for independent samples, and are widely used to validate
random number generators. We use a standard definition of the chi-square
test 
\begin{equation}
\chi_{c}^{2}=\sum_{i=1}^{k}\frac{\left(\bar{P}_{i}-P_{i}^{e}\right)^{2}}{\sigma_{i}^{2}},
\end{equation}
which assumes theoretical and experimental correlations are independent
and approximately Gaussian. The theoretical probability $\bar{P}_{i}$
is obtained by averaging over phase-space ensembles where, for large
ensemble numbers, $\bar{P}_{i}$ converges to the true theoretical
probability $P_{i}$. The experimental probability $P_{i}^{e}$ is
obtained from experimental data or numerically calculated in the case
of exactly known distribution tests. The variance is given by $\sigma_{i}^{2}=\sigma_{e,i}^{2}+\bar{\sigma}_{s,i}^{2}$
and is defined as the sum of the experimental and simulated variances
respectively. For distributions which are approximately Gaussian,
chi-square tests are expected to produce results of $\chi_{c}^{2}/k\approx1$
for $k$ valid bins. We define a valid bin as a bin with more than
$10$ counts. 

\begin{figure}
\begin{centering}
\includegraphics[width=1\columnwidth]{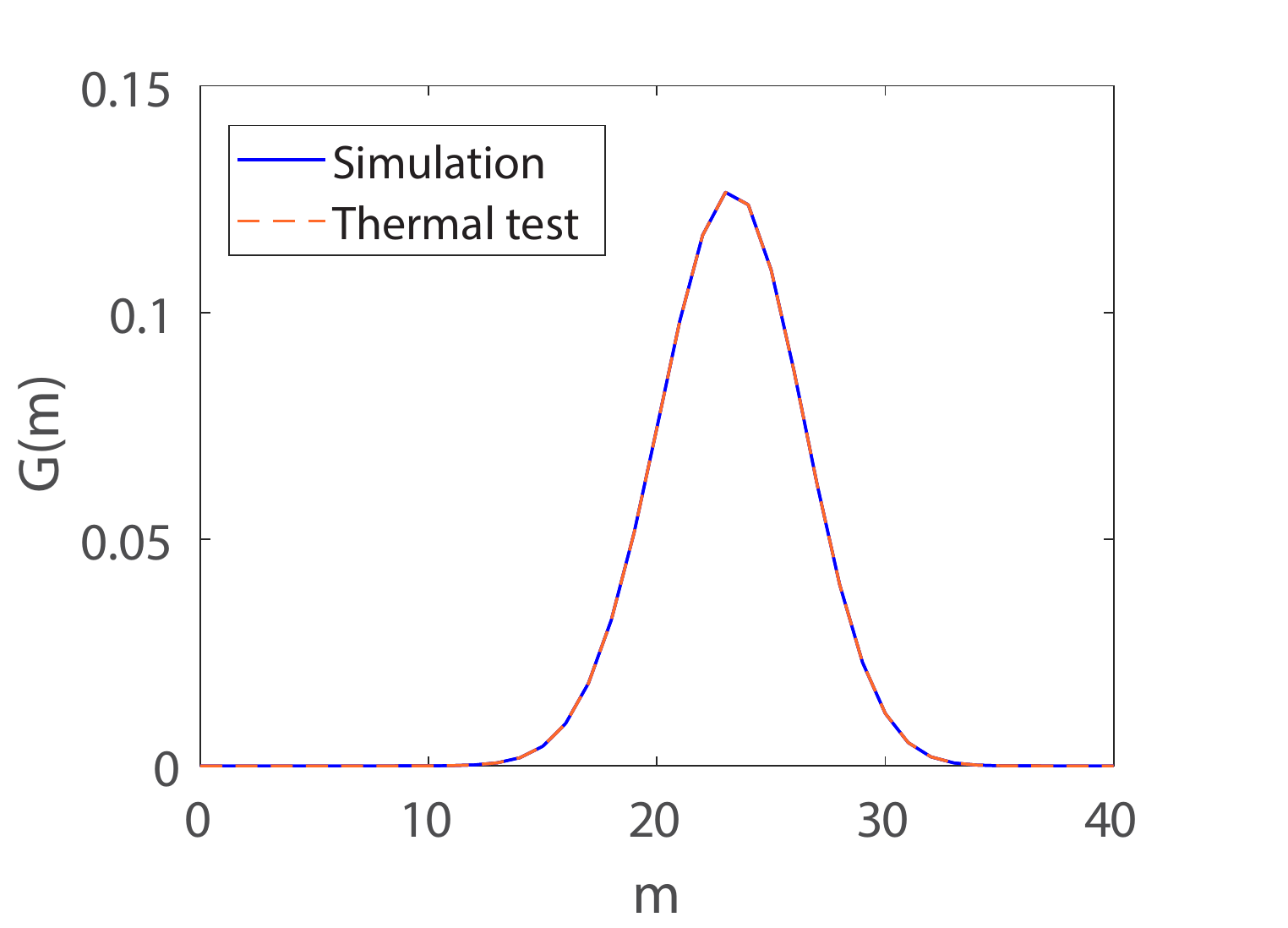}
\par\end{centering}
\caption{Comparison of theory with an exactly known uniform thermal distribution
of total counts $\mathcal{G}_{40}^{(40)}\left(\boldsymbol{m}\right)$
with $1.2\times10^{6}$ ensembles for a $M=40$ GBS network. The solid
blue line is the theoretical prediction, while the orange dash line
is exact distribution. \label{fig:uniform_thermal_total counts}}
\end{figure}

\begin{figure}
\begin{centering}
\includegraphics[width=1\columnwidth]{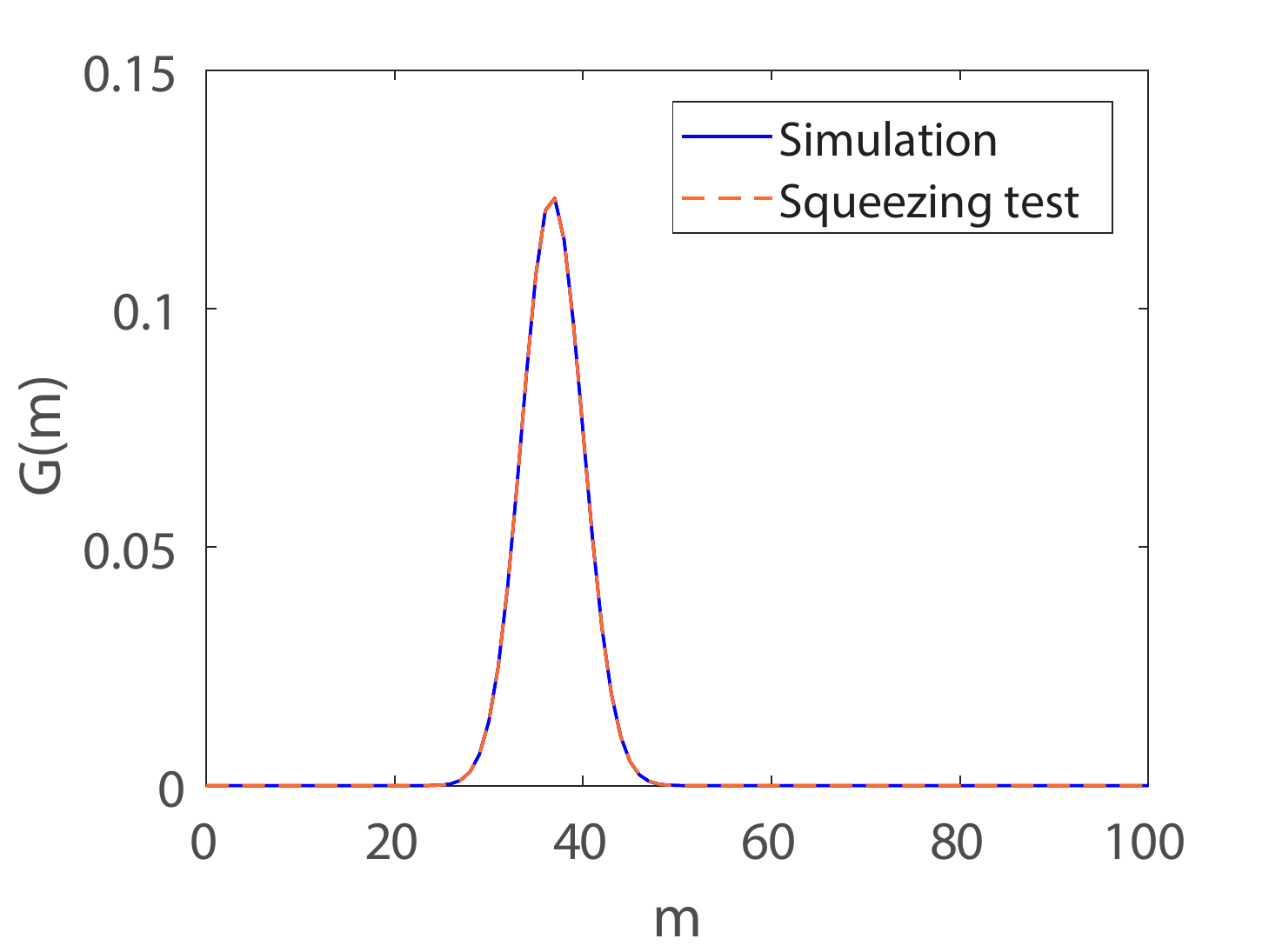}
\par\end{centering}
\caption{Comparison of theory with an exactly known non-uniform squeezed distribution
with $M=100$ mode GBS network. Solid blue line is theoretical prediction
of total counts, $\mathcal{G}_{100}^{(100)}\left(\boldsymbol{m}\right)$,
while the orange dash line is the squeezed distribution. \label{fig:squeezed_total_counts}}
\end{figure}

Before this algorithm was applied to experimental GBS data, it was
thoroughly tested against a variety of exactly known distributions
with a large variety of modes. One such test, is the simulation of
total counts, $\mathcal{G}_{M}^{(M)}\left(\boldsymbol{m}\right)$,
which is the probability of observing $n=M$ counts in any pattern.
Comparisons with an exactly known uniform thermal distribution for
an $M=40$ mode GBS, are shown in Fig. (\ref{fig:uniform_thermal_total counts}).
We see excellent agreement between simulated and thermal correlations
which is validated by a chi-square test result of $\chi_{therm}^{2}/k=1.0\pm0.3$
for $k=32$ valid bins. Comparisons of total counts for a non-uniform
squeezed distribution with $M=100$ modes are also given in Fig. (\ref{fig:squeezed_total_counts}),
which again shows excellent agreement with simulated correlations. 

\begin{figure}
\begin{centering}
\includegraphics[width=1\columnwidth]{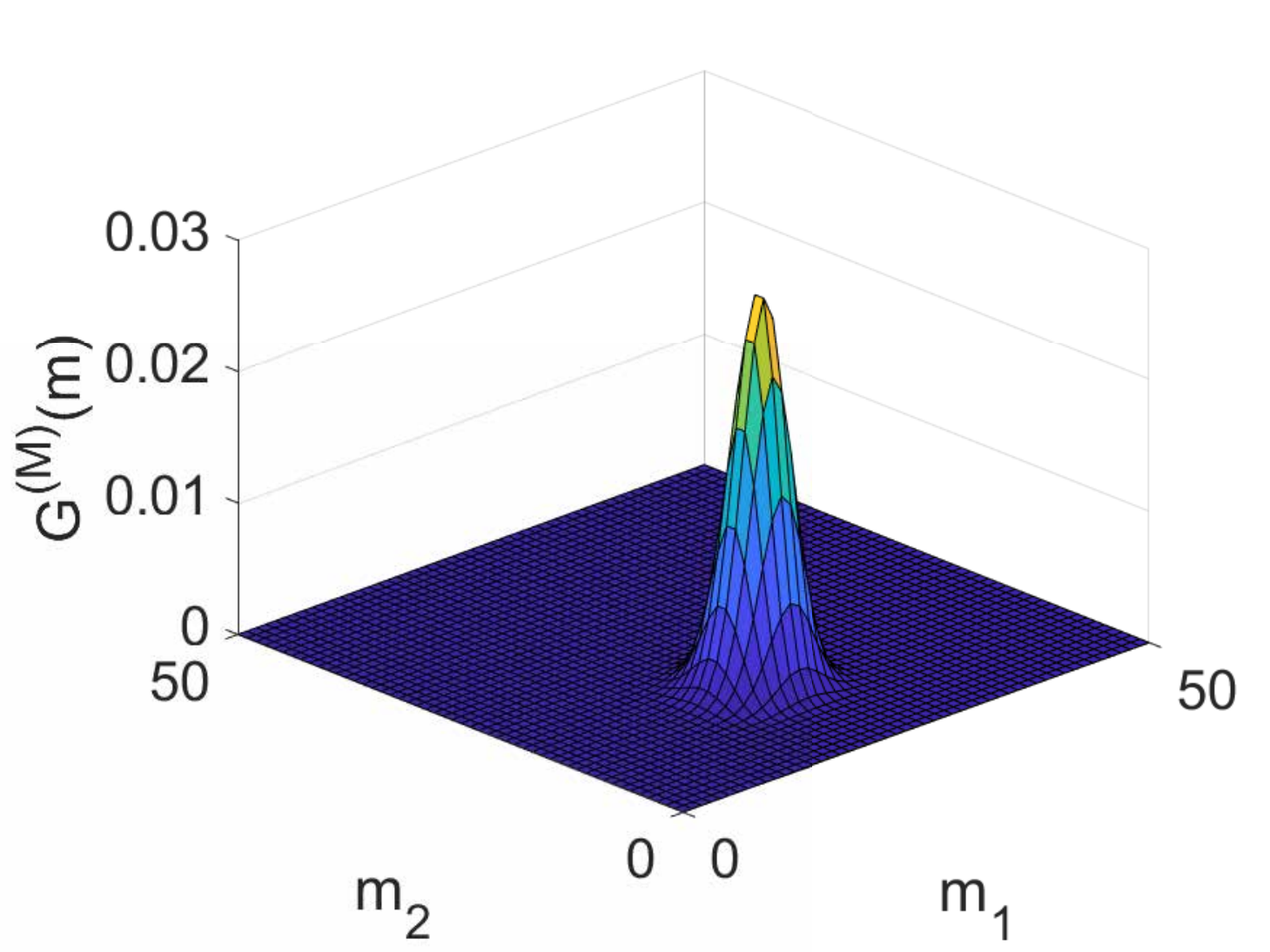}
\par\end{centering}
\caption{Simulation of a non-uniform squeezed distribution for $M=100$ modes
with a two-dimensional binning, $\mathcal{G}_{50,50}^{(100)}\left(\boldsymbol{m}\right)$.
\label{fig:squeezed_two-fold}}

\end{figure}

Another test is shown in Fig. (\ref{fig:squeezed_two-fold}), which
is a two-dimensional binning, , $\mathcal{G}_{M/2,M/2}^{(M)}\left(\boldsymbol{m}\right)$,
of the grouped correlations for the $M=100$ mode non-uniform squeezed
distribution. Chi-square tests were negligible on this scale, being
well within the acceptable margin for an exact distribution. 

The graphs given here are just a small fraction of the possible distributions
one can obtain using grouped correlations. The excellent agreement
of these tests shows how grouped correlations, simulated using stochastic
sampling techniques obtained from the generalized P-representation,
provide a possible answer to the question of verification of linear
bosonic networks. However, a verification algorithm is only useful
if it can simulate correlations in less than exponential time, and
this is where the real benefit of grouped correlations becomes apparent.
The exact phase-space simulations shown here took $<50s$ on a desktop
computer, while $100$-mode GBS simulations took $\sim100s$ \citep{drummond2021simulating}. 

This is orders of magnitude faster than direct simulation methods,
which makes these techniques useful for verification. While up to
$16,000-$th order has been already investigated, these are much more
challenging, due to time and memory constraints. 

\section{N-partite entanglement}

Linear networks generate nonclassical, entangled states. But how does
one verify and classify such complex, high-order entanglement? 

We now consider the generation and detection of an $M$-partite entangled
state using a Gaussian network, expanding on work presented earlier
\citep{drummond2021simulating}. The quadrature phase amplitudes
are defined as $\hat{x}_{j}=\hat{a}_{j}+\hat{a}_{j}^{\dagger}$ and
$\hat{p}_{j}=\frac{1}{i}(\hat{a}_{j}-\hat{a}_{j}^{\dagger})$, in
an appropriate rotating frame. Here, we give an account of how an
$M$-partite entangled state can be generated with just one or two
squeezed input beams, and also discuss the concept of steering for
these multipartite entangled states.

These signatures can all be very easily simulated in phase-space,
due to the direct correspondence of quadrature measurements to Wigner
phase-space simulations for such Gaussian states. We note that this
correspondence is no longer valid for nonlinear networks with small
photon numbers.

\subsection{Two-mode EPR entanglement}

\begin{figure}
\begin{centering}
\includegraphics[width=0.9\columnwidth]{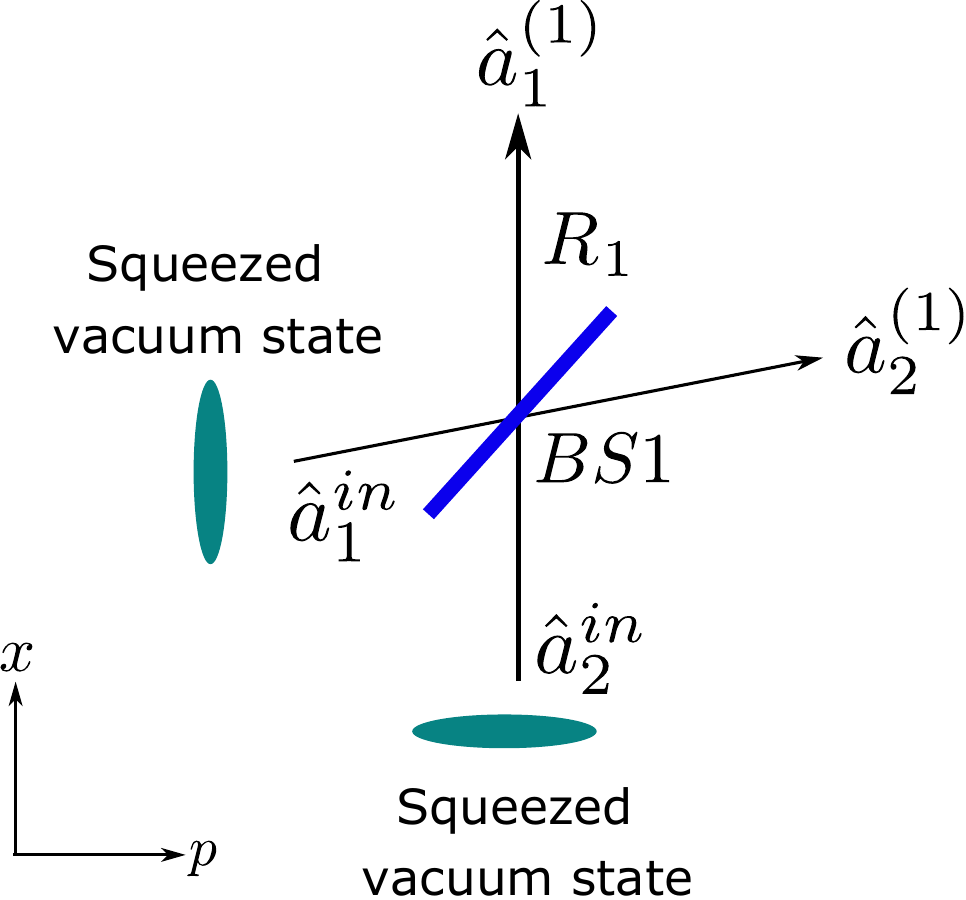}
\par\end{centering}
\caption{Generation of two-mode Einstein-Podolsky-Rosen (EPR) entanglement.
Two squeezed modes $\hat{a}_{1}^{in}$, $\hat{a}_{2}^{in}$ are combined
using a beam splitter labelled $BS1$, that has a reflectivity $R_{1}^{2}$.
The two entangled output modes are $\hat{a}_{1}^{\left(1\right)}$
and $\hat{a}_{2}^{\left(1\right)}$. The first input mode is squeezed
in the $p$-quadrature, i.e. $\Delta^{2}\hat{p}_{1}^{in}=e^{-2r_{1}}$,
while the second input mode is squeezed in the $x$-quadrature with
$\Delta^{2}\hat{x}_{2}^{in}=e^{-2r_{2}}$. As mentioned in the main
text, a choice of $R_{1}=1/\sqrt{2}$, together with these squeezed
inputs, lead to the expressions $\Delta^{2}\left(\hat{x}_{1}^{\left(1\right)}-\hat{x_{2}}^{\left(1\right)}\right)=2e^{-2r_{2}}$
and $\Delta^{2}\left(\hat{p}_{1}^{\left(1\right)}+\hat{p}_{2}^{\left(1\right)}\right)=2e^{-2r_{1}}$,
in (\ref{eq:var-epr}). \label{fig:bipartite_CVEPR}}
\end{figure}

Two-mode Einstein-Podolsky-Rosen (EPR) entanglement is created by
combining one or two squeezed modes across a beam splitter, which
we label $BS1$. This concept was explained for a single squeezed
input in \citet{Reid:1989}. More details are given in \citep{drummond2021simulating,teh2021full}.
The experimental configuration that allows the creation of two-mode
EPR entanglement is given in Fig. \ref{fig:bipartite_CVEPR}.

We consider the unitary transformation represented as
\begin{eqnarray}
\hat{a}_{1}^{(1)} & = & R_{1}\hat{a}_{1}^{in}+T_{1}\hat{a}_{2}^{in}\nonumber \\
\hat{a}_{2}^{(1)} & = & T_{1}\hat{a}_{1}^{in}-R_{1}\hat{a}_{2}^{in}\label{eq:transbs1}
\end{eqnarray}
where $R_{1}^{2}$ is the beam splitter reflectivity and $T_{1}^{2}=1-R_{1}^{2}$
is the transmission coefficient, and we take $R_{1}$ and $T_{1}$
as real and positive. A phase shift may be necessary as well as the
beam splitter to achieve this transformation. The inputs are given
as $\hat{a}_{j}^{in}$. The two outputs of the beam splitter are given
as $\hat{a}_{j}^{(1)}$. For convenience, we denote the variance of
an observable $\text{\ensuremath{\hat{x}}}$ by $\Delta^{2}\hat{x}\equiv(\Delta\hat{x})^{2}=\langle\hat{x}^{2}\rangle-\langle\hat{x}\rangle^{2}$.
Let us suppose $a_{2}^{in}$ is a squeezed vacuum with 
\begin{equation}
\Delta^{2}\hat{x}{}_{2}^{in}=e^{-2r_{2}},
\end{equation}
and $a_{1}^{in}$ is squeezed vacuum with 
\begin{equation}
\Delta^{2}\hat{p}{}_{1}^{in}=e^{-2r_{1}}.
\end{equation}
Here $r_{j}>0$ is the squeezing parameter. We assume minimum uncertainty
states where $(\Delta\hat{x}{}_{j}^{in})(\Delta\hat{p}{}_{j}^{in})=1$.
For $r_{j}=0$, there is no squeezing.

Using (\ref{eq:transbs1}), the variances of the two outgoing fields
can be evaluated. We find on rearranging, 
\begin{eqnarray}
\hat{a}_{1}^{in} & = & R_{1}\hat{a}_{1}^{(1)}+T_{1}\hat{a}_{2}^{(1)}\nonumber \\
\hat{a}_{2}^{in} & = & T_{1}\hat{a}_{1}^{(1)}-R_{1}\hat{a}_{2}^{(1)}.
\end{eqnarray}
Hence, 
\begin{eqnarray}
R_{1}\hat{p}_{1}^{(1)}+T_{1}\hat{p}_{2}^{(1)} & = & \hat{p}_{1}^{in}\nonumber \\
T\hat{x}_{1}^{(1)}-R_{1}\hat{x}{}_{2}^{(1)} & = & \hat{x}_{2}^{in},
\end{eqnarray}
which, for a 50/50 beam splitter where $R_{1}=T_{1}=\frac{1}{\sqrt{2}}$,
leads to $\hat{x}_{1}^{(1)}-\hat{x}{}_{2}^{(1)}=\sqrt{2}\hat{x}_{2}^{in}$
and $\hat{p}_{1}^{(1)}+\hat{p}_{2}^{(1)}=\sqrt{2}\hat{p}_{1}^{in}$.
Therefore,
\begin{eqnarray}
\Delta^{2}(\hat{x}_{1}^{(1)}-\hat{x}_{2}^{(1)}) & = & 2e^{-2r_{2}}\nonumber \\
\Delta^{2}(\hat{p}_{1}^{(1)}+\hat{p}_{2}^{(1)}) & = & 2e^{-2r_{1}}.\label{eq:var-epr}
\end{eqnarray}
For large $r$, both variances become zero. This is the situation
of the Einstein-Podolsky-Rosen (EPR) paradox, where the correlations
between the positions of two separated particles are correlated, and
the momenta of the two particles are anti-correlated \citet{Einstein:1935}.

EPR entanglement can also be created from one squeezed input (letting
$r_{1}=0$) to give 
\begin{eqnarray}
\Delta^{2}(\hat{x}_{1}^{(1)}-\hat{x}_{2}^{(1)}) & = & 2e^{-2r_{2}}\nonumber \\
\Delta^{2}(\hat{p}_{1}^{(1)}+\hat{p}_{2}^{(1)}) & = & 2.\label{eq:one-sq}
\end{eqnarray}
While not the ideal form of EPR entanglement, this reflects an EPR
paradox as we will see below \citep{Reid:1989}.

Entanglement is detected between the two outgoing modes if the inequality
\citep{giovannetti2003characterizing}
\begin{equation}
\Delta(\hat{x}_{1}^{(1)}-g\hat{x}_{2}^{(1)})\Delta(\hat{p}_{1}^{(1)}+h\hat{p}_{2}^{(1)})<1+gh,\label{eq:ent-1}
\end{equation}
where $g$ and $h$ are real positive numbers, is satisfied (take
$g=h=1$) . This implies the two output fields to be entangled for
all values of squeezing, $r_{i}>0$. There is also entanglement for
all $r_{i}>0$ where only one of the input fields is squeezed, as
in (\ref{eq:one-sq}).

\subsection{Two-mode EPR paradox and steering}

It is also possible to examine how to realize an EPR paradox from
the correlations of the two outputs \citep{Einstein:1935}. Here,
we use the approach developed in \citep{Reid:1989}. One considers
how to infer the value for $\hat{x}_{1}^{(1)}$ from a measurement
of $\hat{x}_{2}^{(1)}$; and how to infer the value of $\hat{p}_{1}^{(1)}$
from a measurement of $\hat{p}_{2}^{(1)}$. Alternatively, one may
consider how to infer the value for $\hat{x}_{2}^{(1)}$ from a measurement
of $\hat{x}_{1}^{(1)}$; and how to infer the value of $\hat{p}_{2}^{(1)}$
from a measurement of $\hat{p}_{1}^{(1)}$. The observation of the
paradox is closely connected to the concept of steering. Suppose the
experimenter measures $\hat{x}_{2}^{(1)}$and obtains a value $x_{2}^{(1)}$.
From this value, the prediction $g_{x}x_{2}^{(1)}$ is given for the
outcome of $\hat{x}_{1}^{(1)}$. Similarly, an estimate of $g_{p}p_{2}^{(1)}$
is given for the outcome of $\hat{p}_{1}^{(1)}$, if instead $\hat{p}_{2}^{(1)}$
is measured with result $p_{2}^{(1)}$. Here, $g_{x}$ and $g_{p}$
are real numbers. The variance for the error in the prediction of
$\hat{x}_{1}^{(1)}$ is given as $\Delta^{2}(\hat{x}_{1}^{(1)}-g_{x}\hat{x}_{2}^{(1)})$.

Following previous work \citep{teh2021full}, but allowing for different
values of squeezing in the inputs, we find
\begin{eqnarray}
\Delta^{2}(\hat{x}_{1}^{(1)}-g_{x}\hat{x}_{2}^{(1)}) & = & \Delta^{2}(R_{1}\hat{x}{}_{1}^{in}+T_{1}\hat{x}{}_{2}^{in}-g_{s}(T_{1}\hat{x}_{1}^{in}-R_{1}\hat{x}_{2}^{in}))\nonumber \\
 & = & \Delta^{2}((R_{1}-g_{s}T_{1})\hat{x}{}_{1}^{in}+(T_{1}+g_{s}R_{1})\hat{x}{}_{2}^{in})\nonumber \\
 & = & (R_{1}-g_{s}T_{1})^{2}e^{2r_{1}}+(T_{1}+g_{s}R_{1})^{2}e^{-2r_{2}}
\end{eqnarray}
where we assume the inputs are not correlated. The variance will
be minimum for 
\begin{equation}
g_{x}=\frac{R_{1}T_{1}(e^{2r_{1}}-e^{-2r_{2}})}{(T_{1}^{2}e^{2r_{1}}+R_{1}^{2}e^{-2r_{2}})}.\label{eq:g_xs}
\end{equation}
Similarly, 
\begin{eqnarray}
\Delta^{2}(\hat{p}_{1}^{(1)}+g_{p}\hat{p}{}_{2}^{(1)}) & = & g_{p}^{2}(T_{1}^{2}e^{-2r_{1}}+R_{1}^{2}e^{2r_{2}})\nonumber \\
 &  & +R_{1}^{2}e^{-2r_{1}}+T_{1}^{2}e^{2r_{2}}\nonumber \\
 &  & -2R_{1}T_{1}g_{p}(e^{2r_{2}}-e^{-2r_{1}}),
\end{eqnarray}
which is minimized for 
\begin{equation}
g_{p}=\frac{R_{1}T_{1}(e^{2r_{2}}-e^{-2r_{1}})}{(R_{1}^{2}e^{2r_{2}}+T_{1}^{2}e^{-2r_{1}})}.\label{eq:g_ps}
\end{equation}
This gives on substitution
\begin{eqnarray}
\Delta^{2}(\hat{x}_{1}^{(1)}-g_{x}\hat{x}_{2}^{(1)}) & = & \frac{e^{-2r_{2}}}{\left(1-R_{1}^{2}\right)+R_{1}^{2}e^{-2(r_{2}+r_{1})}}\\
\Delta^{2}(\hat{p}_{1}^{(1)}+g_{p}\hat{p}{}_{2}^{(1)}) & = & \frac{e^{-2r_{1}}}{R_{1}^{2}+\left(1-R_{1}^{2}\right)e^{-2(r_{1}+r_{2})}}.\label{eq:varsteer-x}
\end{eqnarray}
The inequality
\begin{equation}
S_{1|2}\equiv\Delta(\hat{x}_{1}^{(1)}-g_{x}\hat{x}_{2}^{(1)})\Delta(\hat{p}_{1}^{(1)}+g_{p}\hat{p}{}_{2}^{(1)})<1\label{eq:epr-crit}
\end{equation}
gives the condition for the EPR paradox \citep{Reid:1989}, and is
sufficient to confirm an EPR steering of the mode $1$ by measurements
on mode $2$ \citep{Cavalcanti2009,Wiseman_PRL2007}. 

The right-hand side of the inequality is determined by the minimum
value of the uncertainty product for $\hat{x}_{1}$ and $\hat{p}_{1}$:
$(\Delta\hat{x}{}_{1})(\Delta\hat{p}{}_{1})=1$. The steering party,
mode $2$, is not relevant to the bound of the inequality. When the
inequality is satisfied, the errors in the inferences for the values
of $\hat{x}_{1}$ and $\hat{p}_{1}$ are sufficiently small that they
cannot be explained consistently with the notion of a local quantum
state for the system $1$ and the concept of local realism as defined
by EPR.

Provided $0<R_{1}<1$, there is a perfect EPR correlation as the variances
become zero, as one or both of $r_{j}\rightarrow\infty$. One may
prove by differentiation with respect to $R=R_{1}^{2}$ that the optimal
choice to minimize the steering product is $R_{1}=1/\sqrt{2}$. The
value of $\frac{dS_{1|2}}{dR}$ is a fraction with a positive denominator
and a numerator given by $-e^{-2r_{s}}(1-e^{-2r_{s}})^{2}\left(1-2R\right)$,
where $r_{s}=r_{1}+r_{2}$, which indicates $R=1/2$ to be a minimum
for all values of $r_{s}$. For $R_{1}^{2}=1/2$, the optimal steering
product becomes 
\begin{equation}
S_{1|2}=\frac{1}{\cosh(r_{1}+r_{2})}\label{eq:epr-solnS}
\end{equation}
as defined for the optimal gains 
\begin{equation}
g_{x,s}=-g_{p,s}=\frac{(e^{2r_{1}}-e^{-2r_{2}})}{(e^{2r_{1}}+e^{-2r_{2}})}\,.\label{eq:gains-epr-soln}
\end{equation}

In this work, it is shown how the steering can be detected, as well
as the entanglement, for all values of $r_{j}>0$. Steerable states
are useful because they allow a one-sided device-independent detection
of entanglement \citep{Branciard_PRA2012,Opanchuk_PRA2014,Uola2020_RMP}.
This is evident on noticing that the bound on the right side of the
inequality (\ref{eq:epr-crit}) giving the EPR criterion arises only
from the uncertainty product of mode $1$. The detection would not
involve a calibration of the quantum noise level of the mode $2$.
This also explains why the level of correlation required to detect
the steering as opposed to the entanglement is greater. The bound
of the entanglement condition (\ref{eq:ent-1}) involves a calibration
of the quantum noise level of both modes.

While it seems that steering and entanglement can both be detected
for all $r>0$, the above calculations do not account for decoherence
since we have assumed pure states. The impact of decoherence on steering
as opposed to entanglement is greater and has been studied in many
works (see for example \citep{rosales2015decoherence,Reid:2009_RMP81}).
In particular, there is an asymmetrical effect depending on which
mode is exposed to the decoherence e.g. for thermal noise \citep{he2013einstein,KiesewetterPhysRevA_2014}.

\subsection{$M$-partite entanglement and steering}

\begin{figure}
\begin{centering}
\includegraphics[width=0.9\columnwidth]{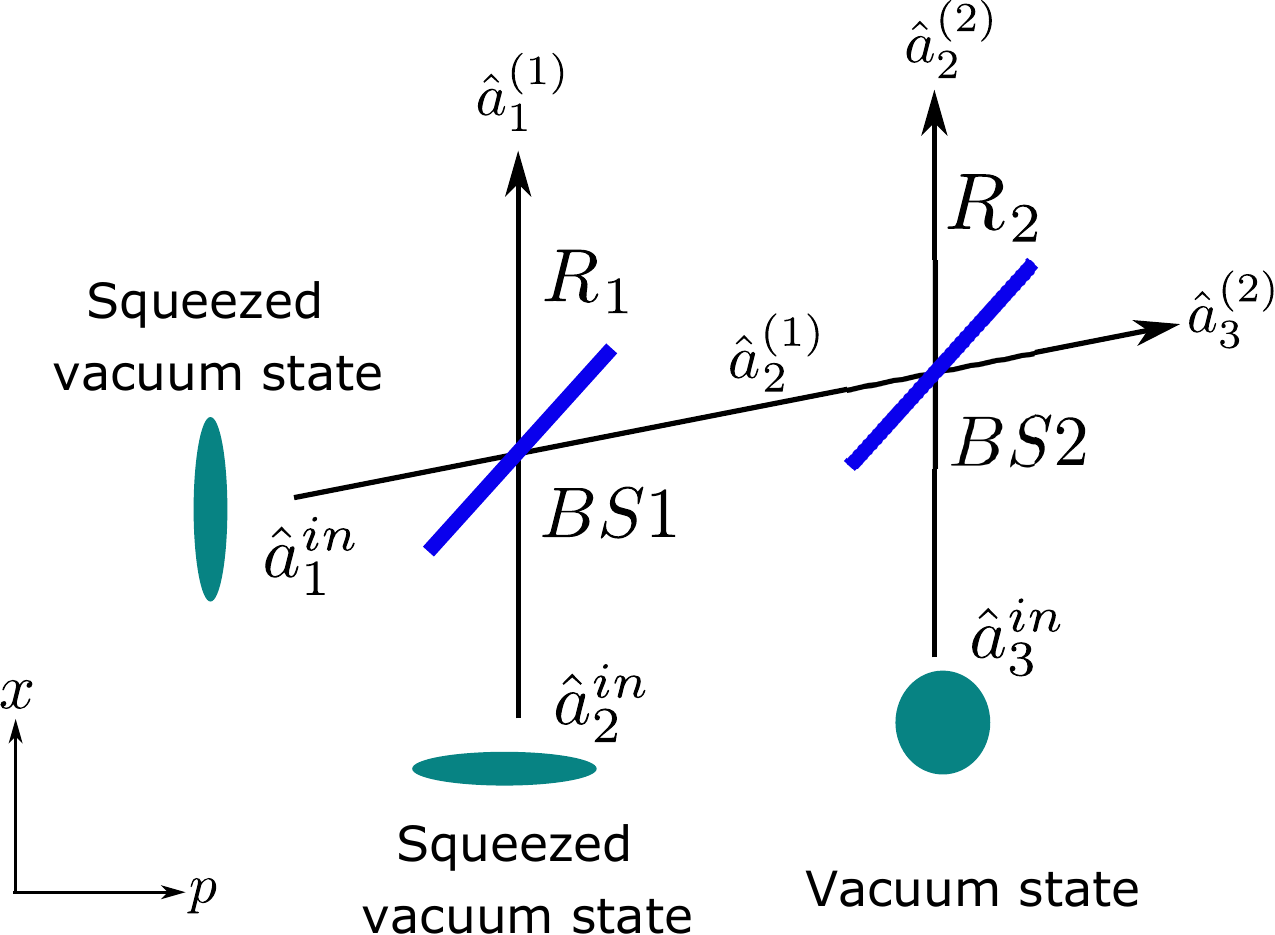}
\par\end{centering}
\caption{Generation of tripartite entanglement. Two squeezed modes $\hat{a}_{1}^{in}$,
$\hat{a}_{2}^{in}$ are first combined using a beam splitter labelled
$BS1$ that has a reflectivity $R_{1}^{2}$. One output of $BS1$
is subsequently combined with a third input $\hat{a}_{3}^{in}$ using
a second beam splitter $BS2$. The three entangled output modes are
$\hat{a}_{1}^{\left(1\right)}$, $\hat{a}_{2}^{\left(2\right)}$ and
$\hat{a}_{3}^{\left(2\right)}$. The first input mode is squeezed
in the $p$-quadrature, i.e. $\Delta^{2}\hat{p}_{1}^{in}=e^{-2r_{1}}$,
while the second input mode is squeezed in the $x$-quadrature with
$\Delta^{2}\hat{x}_{2}^{in}=e^{-2r_{2}}$, and the third input mode
is a vacuum state. The generalization to $M$-partite entanglement
is achieved by simply extending the above experimental configuration
using $M-1$ beam splitters, as described in the main text. \label{fig:tripartite_CVEPR}}
\end{figure}

To generate multipartite entanglement, more modes are needed. To generate
tripartite entanglement, the field $a_{2}^{(1)}$ is combined with
a vacuum state of a third mode using a second beam splitter $BS2$,
with reflectivity $R_{2}^{2}$ and $T_{2}^{2}=1-R_{2}^{2}$. Here,
we follow the method introduced by van Loock and Furusawa \citep{van2003detecting}
and extended by \citep{teh2014criteria}. Fig. \ref{fig:tripartite_CVEPR}
gives a schematic of the configuration for tripartite entanglement
generation. The transformation is
\begin{eqnarray}
\hat{a}_{2}^{(2)} & = & R_{2}\hat{a}_{2}^{(1)}+T_{2}\hat{a}_{3}^{in}\nonumber \\
\hat{a}_{3}^{(2)} & = & T_{2}\hat{a}{}_{2}^{(1)}-R_{2}\hat{a}_{3}^{in}.
\end{eqnarray}
The output of field $a_{2}$ is denoted $\hat{a}_{2}^{(2)}$. For
$M=3$, there are only two beam splitters, and the output of $a_{3}$
is $\hat{a}_{3}^{(2)}$. Rearranging, we find $\hat{a}_{2}^{(1)}=R_{2}\hat{a}_{2}^{(2)}+T_{2}\hat{a}_{3}^{(2)}$.
Thus 
\begin{eqnarray}
\hat{x}_{2}^{(1)} & = & R_{2}\hat{x}_{2}^{(2)}+T_{2}\hat{x}_{3}^{(2)}\nonumber \\
\hat{p}_{2}^{(1)} & = & R_{2}\hat{p}_{2}^{(2)}+T_{2}\hat{p}_{3}^{(2)}.\label{eq:bs2x,p}
\end{eqnarray}
Taking $R_{1}=R_{2}=\frac{1}{\sqrt{2}}$, from (\ref{eq:var-epr})
we see that $\Delta^{2}(\hat{x}_{1}^{(1)}-\frac{1}{\sqrt{2}}(\hat{x}_{2}^{(2)}+\hat{x}_{3}^{(2)}))=2e^{-2r_{2}}$
and $\Delta^{2}(\hat{p}_{1}^{(1)}+\frac{1}{\sqrt{2}}(\hat{p}_{2}^{(2)}+\hat{p}_{3}^{(2)}))=2e^{-2r_{1}}$.
These variances are both zero for large $r_{i}$, and the three output
fields satisfy the condition
\begin{equation}
\Delta(\hat{x}_{1}^{(1)}-\frac{1}{\sqrt{2}}(\hat{x}_{2}^{(2)}+\hat{x}_{3}^{(2)}))\Delta(\hat{p}_{1}^{(1)}+\frac{1}{\sqrt{2}}(\hat{p}_{2}^{(2)}+\hat{p}_{3}^{(2)}))<1\label{eq:cond-ent3}
\end{equation}
known to certify genuine tripartite entanglement. This condition was
proved to be sufficient for the verification of full tripartite inseparability
in \citep{van2003detecting}, and was proved sufficient to confirm
genuine tripartite entanglement in \citep{teh2014criteria}. The reader
is referred to equation (17) of \citep{teh2014criteria} with equal
gains for the second and third modes. In fact, the product on the
left side of the inequality becomes zero, indicating a maximum EPR
entanglement, in the limit of large $r_{i}$. 

The EPR paradox is seen to hold for $r_{1}$ and $r_{2}$ becoming
large. The EPR correlations of the two-mode state are transferred
directly onto the three modes. The paradox is observed because measurements
can be made on the two fields $a_{2}$ and $a_{3}$ to give the value
of $\frac{1}{\sqrt{2}}(\hat{x}_{2}^{(2)}+\hat{x}_{3}^{(2)})$, from
which the outcome of $\hat{x}_{1}^{(1)}$ can be predicted with certainty.
Similarly, the outcome of $\hat{p}_{1}^{(1)}$ can be predicted with
certainty by measurement of $\frac{1}{\sqrt{2}}(\hat{p}_{2}^{(2)}+\hat{p}_{3}^{(2)})$.
When the fields are spatially separated, the premise of local realism
can be applied and the paradox follows. In fact, the condition (\ref{eq:cond-ent3})
is seen to be the condition given by (\ref{eq:epr-crit}) for an EPR
paradox. The condition is also a verification of the EPR steering
of mode $1$, by measurements made on modes $2$ and $3$.

This leaves us to examine the EPR steering condition more carefully.
On taking $R_{2}=\frac{1}{\sqrt{2}}$, from (\ref{eq:varsteer-x})
we see on substituting (\ref{eq:bs2x,p}), 
\begin{eqnarray}
\Delta^{2}(\hat{x}_{1}^{(1)}-\frac{g_{x}}{\sqrt{2}}(\hat{x}_{2}^{(2)}+\hat{x}_{3}^{(2)})) & = & \frac{2e^{-2r_{2}}}{1+e^{-2(r_{2}+r_{1})}}\nonumber \\
\Delta^{2}(\hat{p}_{1}^{(1)}+\frac{g_{p}}{\sqrt{2}}(\hat{p}_{2}^{(2)}+\hat{p}_{3}^{(2)})) & = & \frac{2e^{-2r_{1}}}{1+e^{-2(r_{1}+r_{2})}}\,.\label{eq:steer-var}
\end{eqnarray}
Hence the steering product defined for the three output modes
is
\begin{equation}
S_{1|23}\equiv\Delta(\hat{x}_{1}^{(1)}-\frac{1}{\sqrt{2}}(\hat{x}_{2}^{(2)}+\hat{x}_{3}^{(2)}))\Delta(\hat{p}_{1}^{(1)}+\frac{1}{\sqrt{2}}(\hat{p}_{2}^{(2)}+\hat{p}_{3}^{(2)}))\label{eq:steer3}
\end{equation}
which becomes 
\begin{equation}
S_{1|23}=\frac{1}{\cosh(r_{1}+r_{2})}
\end{equation}
as given by (\ref{eq:epr-solnS}). There is steering of system $1$
for $r_{1}>0$ or $r_{2}>0$, or both. A single squeezed input is
all that is required, in principle, although stronger steering is
obtained for smaller variances which are better achieved with two
squeezed input fields. Properties of genuine multipartite steering
are studied elsewhere \citep{teh2021full}.

To generate four-partite entanglement where $M=4$, the process continues
with another beam splitter, $BS3$.
\begin{eqnarray}
\hat{a}_{3}^{(3)} & = & R_{3}\hat{a}_{3}^{(2)}+T_{3}\hat{a}_{4}^{in}\nonumber \\
\hat{a}_{4}^{(3)} & = & T_{3}\hat{a}{}_{3}^{(2)}-R_{3}\hat{a}_{4}^{in}.
\end{eqnarray}
The output of mode $\hat{a}_{3}$ is denoted $\hat{a}_{3}^{(3)}$
and the output of mode $\hat{a}_{4}$ is denoted $\hat{a}_{4}^{(3)}$.
For a suitable choice of $R_{3}$, the four outputs $a_{1}^{(1)}$,
$a_{2}^{(2)}$, $a_{3}^{(3)}$and $a_{4}^{(3)}$ are genuinely $4$-partite
entangled.

In fact, it is possible to continue the process of splitting the
final field using a beam splitter, to create a system of $M$ output
modes from a string of $M-1$ beam splitters. The reflectivities of
the string of beam splitters can be selected so that the correlations
are governed by the two-mode EPR correlations emerging from the first
two beam splitters. For the $M$ modes given by outputs $a_{1}^{(1)}$,
$a_{2}^{(2)}$, $a_{3}^{(3)}$, $..,a_{M-1}^{(M-1)}$, $a_{M}^{(M-1)}$,
it follows that 
\begin{align}
\Delta^{2}(\hat{x}_{1}^{(1)}-\frac{1}{\sqrt{M-1}}(\hat{x}_{2}^{(2)}+\hat{x}_{3}^{(3)}+..\hat{x}_{M}^{(M-1)})) & =2e^{-2r_{2}}\nonumber \\
\Delta^{2}(\hat{p}_{1}+\frac{1}{\sqrt{M-1}}(\hat{p}_{2}^{(2)}+\hat{p}_{3}^{(3)}+..\hat{p}_{M}^{(M-1)})) & =2e^{-2r_{1}},
\end{align}
which gives genuine $M$-partite entanglement. This is done by selecting
for the beam splitters, $R_{M-1}^{2}=1/2$, $R_{M-2}^{2}=\frac{1}{3}$,
$R_{M-j}^{2}=\frac{1}{j+1}$ for $j<M-1$, with $R_{1}^{2}=1/2$ \citep{van2003detecting,teh2014criteria}.
A criterion to confirm the detection of the genuine $M$-partite entanglement
is defined in the next section.\\
\begin{figure}
\begin{centering}
\includegraphics[width=0.9\columnwidth]{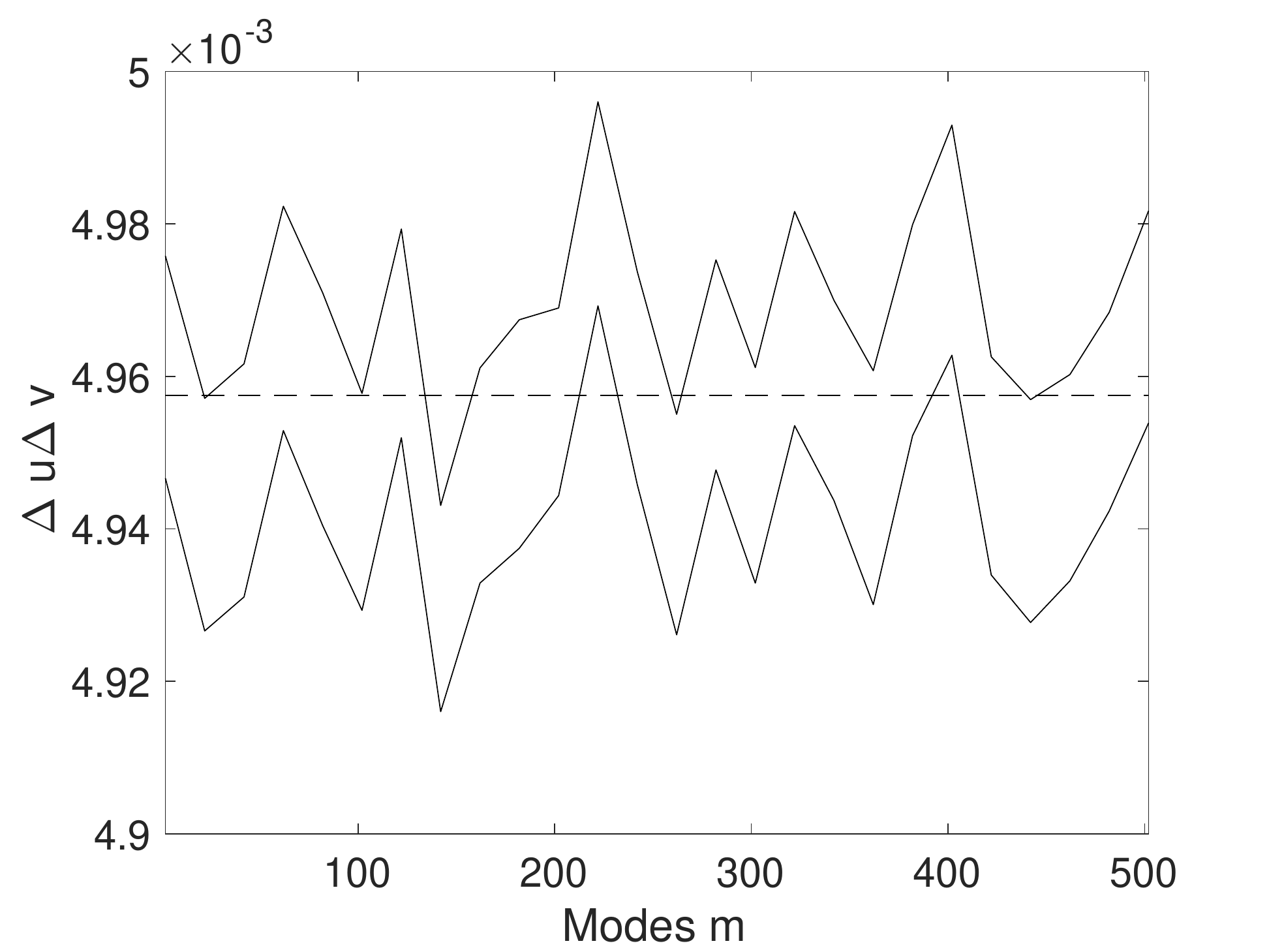}
\par\end{centering}
\caption{Simulation of multipartite entanglement in the Wigner representation,
with two $r=3$ squeezed inputs. Sampling errors are indicated by
the upper and lower error bars. Exact results are shown by the dashed
line. In these simulations, there were 120 repeats of a 1000 sample
parallel ensemble, to obtain sampling error estimates. Sampling errors
are independent of the mode number, at about $\pm10^{-5}.$ \label{fig:Wigner-product}}
\end{figure}

\begin{figure}
\begin{centering}
\includegraphics[width=0.9\columnwidth]{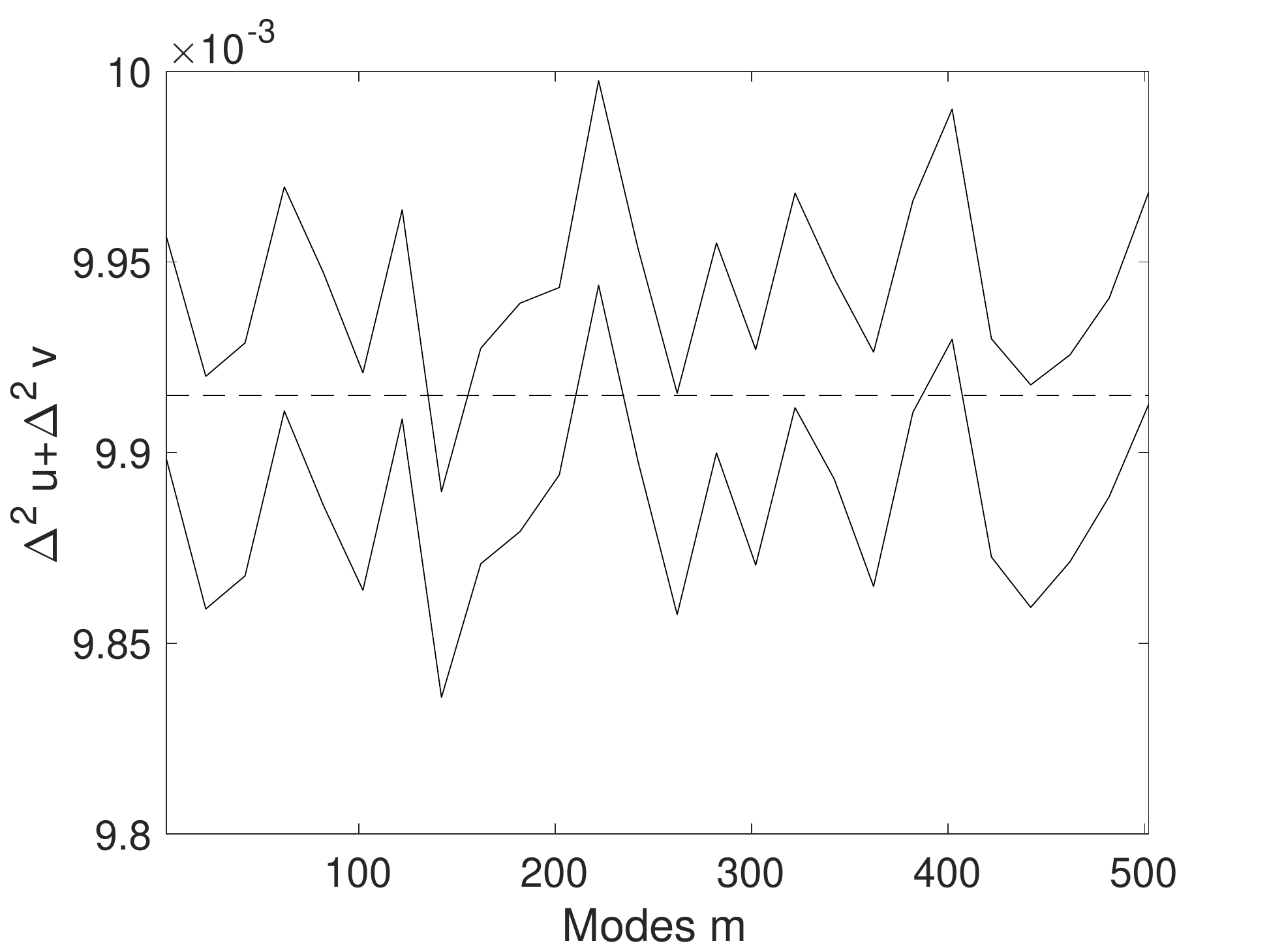}
\par\end{centering}
\caption{Simulation of multipartite entanglement in the Wigner representation,
with two $r=3$ squeezed inputs. Sampling errors are indicated by
the upper and lower error bars. Exact results are shown by the dashed
line. In these simulations, there were 120 repeats of a 1000 sample
parallel ensemble, to obtain sampling error estimates. Sampling errors
are independent of the mode number, at about $\pm2\times10^{-5}.$
\label{fig:Wigner-sum}}
\end{figure}

\subsection{Detecting M-partite entanglement}

Let $\hat{u}=\hat{x}_{1}-\frac{1}{\sqrt{M-1}}(\hat{x}_{2}+\hat{x}_{3}+..\hat{x}_{M})$
and $\hat{v}=\hat{p}_{1}+\frac{1}{\sqrt{M-1}}(\hat{p}_{2}+\hat{p}_{3}+..\hat{p}_{M})$
be the quadrature amplitudes as defined in the section above. Then
the observation of 
\begin{equation}
(\Delta\hat{u})(\Delta\hat{v})<\frac{2}{\left(M-1\right)}
\end{equation}
confirms $M$-partite entanglement for all $M$. The observation of
\begin{equation}
(\Delta\hat{u})^{2}+(\Delta\hat{v})^{2}<\frac{4}{M-1}
\end{equation}
also confirms $M$-partite entanglement for all $M$. The second
result follows from the first, because if $\Delta^{2}u+\Delta^{2}v<\frac{4}{(N-1)}$,
then it must be true that $\Delta u\Delta v<\frac{2}{M-1}$ (use $x^{2}+y^{2}\geq2xy$).
The second result was derived in \citep{van2003detecting} for the
confirmation of full $M$-partite inseparability. Typical simulation
results are shown in Figs (\ref{fig:Wigner-product}) and (\ref{fig:Wigner-sum}),
for $r=3$ inputs, showing complete agreement with analytic theory,
up to the sampling error. Clearly, multimode entanglement is only
confirmed for the product method at large $M$.

Clearly, the outputs of the above network satisfy this criterion for
large enough $r_{i}$, with $\Delta u\Delta v=2e^{-(r_{1}+r_{2})}<\frac{2}{M-1}$.
With $r_{1}=r_{2}$, we can detect the entanglement using this criterion
for squeeze parameter $r>\frac{1}{2}ln(M-1)$. We mention that the
criterion is sufficient but not necessary to detect the multipartite
entanglement, and other methods are possible. This is especially true
if one is justified to assume pure or Gaussian states, or is interested
in determining only full multipartite inseparability. The reader is
referred to \citep{sperling2013multipartite,gerke2015full,Villar_PRL2005,Chen2014PhysRevLett.112.120505,coelho2009three,shalm2013three}.
The above criterion however does not require these assumptions. The
use of just one squeezed input will also generate an $M$-partite
entangled state, for larger values of $r$. The use of more squeezed
inputs allows generation of CV GHZ and cluster states that have an
improved scaling for $M$ \citep{van2003detecting}.

Proof: The proof follows that given in \citep{van2003detecting},
but is modified to account for a product criterion, and to allow for
mixed states. We follow the methods developed in \citep{teh2014criteria}.
Consider constants $g$ and $h$: Then for mixtures, it is always
true \citep{teh2021full} that
\begin{eqnarray}
\Delta(u)\Delta(v) & \geq & \sum_{R}P_{R}\Delta_{R}u\Delta_{R}v
\end{eqnarray}
where $u=x_{1}-g(x_{2}+x_{3}+...+x_{M})$ and $v=p_{1}+h(p_{2}+p_{3}+..+p_{M})$.
Here we assume $\rho=\sum_{R}P_{R}\rho_{R}$ where $\rho_{R}$ is
a pure state, and $\sum_{R}P_{R}=1$ where $P_{R}>0$ is a probability.
The $(\Delta_{R}u)^{2}$ denotes the variance of $u$ with respect
to the pure state $R$.

To prove $M$-partite entanglement, one needs to prove entanglement
across every bipartition. Consider a pure product state that is separable
along bipartition $R-S$, where $R$ contains $M_{R}$ systems and
$S$ contains $M_{S}$ systems, such that $M_{R}+M_{S}=M$. Then let
$R$ contain system $1$, and we note that $M_{S}\geq1$ and $M_{R}\geq1$.
Let the modes $1$, $i,..$ be elements of $R$ and the modes $j$,
$k$,.. be elements of $S$. For a product state, as arises from pure
separable systems $A$ and $B$, the variance of the sum of two observables
$X_{A}+X_{B}$ will satisfy 
\begin{eqnarray}
(\Delta(X_{A}+X_{B}))^{2} & = & (\Delta X_{A})^{2}+(\Delta X_{B})^{2}
\end{eqnarray}
where here $X_{A}$ ($X_{B}$) is an observable for system $A$ ($B$).
Therefore,
\begin{eqnarray}
(\Delta u\Delta v)^{2} & = & \{(\Delta(x_{1}-gx_{i}+..)^{2}+(\Delta(gx_{j}+gx_{k}+..)^{2}\}\nonumber \\
 &  & \times\{(\Delta(p_{1}+gp_{i}+.))^{2}+(\Delta(gp_{j}+gp_{k}+..)^{2}\}\nonumber \\
 & \geq & \{\Delta(x_{1}-gx_{i}+..)(\Delta(p_{1}+gp_{i}+..)\nonumber \\
 &  & +(\Delta(gx_{j}+gx_{k}+..)(\Delta(gp_{j}+gp_{k}+..)\}^{2}\nonumber \\
 & \geq & (\frac{M_{s}}{M-1}+\frac{M_{s}}{M-1})^{2}\nonumber \\
 & \geq & \frac{4}{(M-1)^{2}}
\end{eqnarray}
where we have used $(a^{2}+b^{2})(c^{2}+d^{2})\geq(|ac|+|bd|)^{2}$,
which is easily proved on noting that $(bc)^{2}+(ad)^{2}\geq2|acbd|$.
We have also  used 
\begin{eqnarray}
\Pi_{1} & \equiv & \Delta(x_{1}-gx_{i}+..)\Delta(p_{1}+gp_{i}+..)\nonumber \\
 & \geq & \frac{1}{2}|\langle[x_{1}-gx_{i}+..,p_{1}+hp_{i}+..]\rangle|\\
 & \geq & \frac{1}{2}|\langle[x_{1},p_{1}]-gh[x_{i},p_{i}]+..\rangle|\nonumber \\
 & = & 1-(M_{R}-1)gh\nonumber \\
 & = & 1-\frac{(M_{R}-1)}{M-1}\nonumber \\
 & = & \frac{M-M_{R}}{M-1}=\frac{M_{s}}{M-1}
\end{eqnarray}
and, similarly,
\begin{eqnarray}
\Pi_{j} & \equiv & \Delta(x_{j}+x_{k}+..)\Delta(p_{j}+p_{k}+..)\nonumber \\
 & \geq & \frac{1}{2}|\langle[x_{j}+x_{k}+..,p_{j}+p_{k}+..]\rangle|\\
 & \geq & \frac{1}{2}|\langle[x_{j},p_{j}]+[x_{k},p_{k}]+..\rangle|\nonumber \\
 & = & M_{S}\,.
\end{eqnarray}
This gives the required entanglement result, with a typical simulation
result in Fig (\ref{fig:Wigner-product}), using $r=3$ squeezed inputs.
$\square$ 

\begin{figure}
\begin{centering}
\includegraphics[width=0.9\columnwidth]{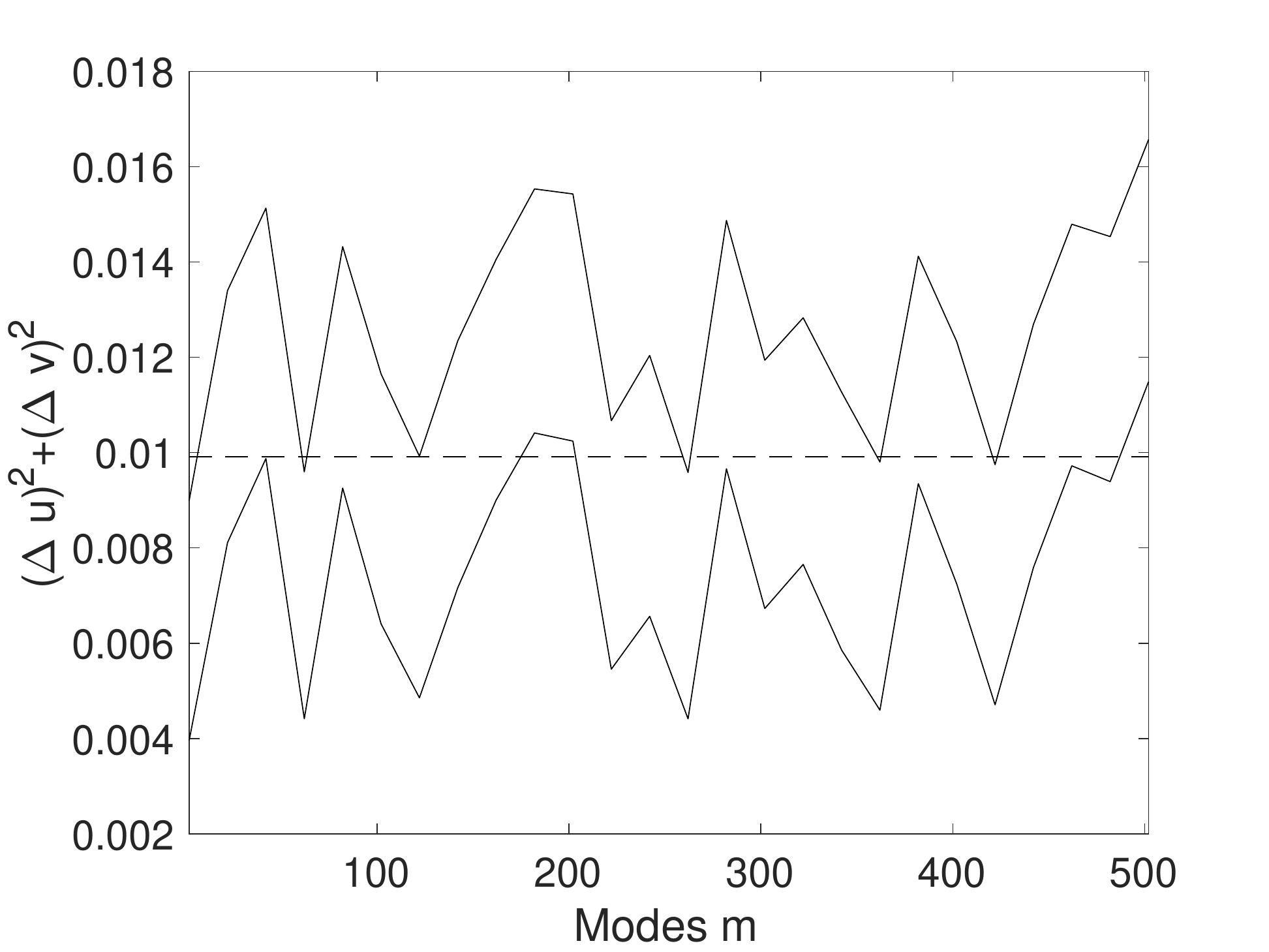}
\par\end{centering}
\caption{Simulation of multipartite entanglement in the positive-P representation,
with two $r=3$ squeezed inputs. Sampling errors are indicated by
the upper and lower error bars. Exact results are shown by the dashed
line. In these simulations, there were 1200 repeats of a 2000 sample
parallel ensemble, to obtain sampling error estimates. Sampling errors
are independent of the mode number, at about $\pm2\times10^{-3}.$
Despite the $20\times$ larger ensemble, sampling errors are still
two orders of magnitude larger than for the Wigner simulations given
above. \label{fig:+P-sum}}
\end{figure}

All of these rather complex operations can be readily simulated using
a simple unitary transformation on the amplitudes, and typical results
for up to $500\times500$ unitaries are given in Figs (\ref{fig:Wigner-product})
and (\ref{fig:Wigner-sum}). However, in this case it is the Wigner
representation that is the most efficient simulation technique, not
the positive-P representation. This has a much larger sampling error,
as shown in Fig (\ref{fig:+P-sum}). The reason for this is that the
Wigner representation directly represents the symmetric operator ordering
of the quadrature measurements. When there is a mismatch between the
measurement ordering and the representation ordering, corrections
are needed that increase the sampling error.

\section{Outlook}

We have given a short tutorial account of recent developments in the
use of phase-space simulations to treat large, linear networks. These
methods can efficiently predict the moments and correlations of Gaussian
boson samplers to all orders. This allows an assessment of whether
such devices are in fact performing as expected theoretically. The
addition of extra noise and decoherence as found experimentally can
be easily included. In addition, it is possible to simulate the detection
of a large variety of other nonclassical signatures, including multipartite
EPR effects.

The focus here is on classical simulations that can verify measured
correlations and probability distributions in linear networks. We
find similar levels of complexity between the experiments and the
verification simulations, meaning that the simulations themselves
are reasonably fast. They are mostly limited by the Fourier transforms
needed to bin the data in the GBS case. 

However, this does not mean that the classical computer has really
performed the same task in a similar time to the quantum experiment.
Actually, for the generation of entangled states this is true, as
the outputs are identical. For the GBS case, which is known to be
a complex task, the simulations generate random complex numbers that
allow verification of the GBS experimental data after binning, but
they do not directly replicate the experimentally generated bits. 

We also showed that while the positive-P method is well-suited to
photon-counting and GBS verification, the Wigner method has lower
sampling errors for simulating quadrature measurements, and multipartite
entanglement. Nonlinear effects were not treated here. However, these
are known to occur in such optical devices. It is not impossible to
include them also, as previously demonstrated in nonlinear fiber-optical
\citep{DrummondGardinerWalls1981,carter1987squeezing} and BEC \citep{Deuar:2007_BECCollisions}
systems. Nonlinearities are detrimental in the GBS case, but they
are essential in other types of device such as the CIM \citep{kiesewetter2021weighted}.

While a tutorial of this type cannot cover all the recent developments,
we also note two recent advances in theory \citep{villalonga2021efficient}
and experiment \citep{zhong2021phase}.

\subsubsection*{Acknowledgments}

This work was funded through the Australian Research Council Discovery
Project scheme under Grants DP180102470 and DP190101480, and through
a research grant from NTT Research Corporation.

\bibliographystyle{apsrev4-2}

\end{document}